\documentclass[3p,12pt]{elsarticle}
\usepackage{booktabs}
\usepackage[fleqn]{amsmath}
\usepackage{cases}
\usepackage{multirow}
\usepackage{xcolor}
\usepackage[normalem]{ulem}
\usepackage{tabularx}
\usepackage{siunitx}
\usepackage{comment}
\usepackage{algorithm}
\usepackage{algpseudocode}
\usepackage{algorithmicx}
\usepackage{grffile}
\usepackage{rotating}
\usepackage{lineno}
\usepackage{hyperref}
\usepackage{graphicx,enumerate,subfigure}
\usepackage{amssymb}
\usepackage{bm,amsmath}
\usepackage{mathtools}
\usepackage{caption,color,xcolor}
\usepackage{subeqnarray}
\usepackage{multirow}
\usepackage{lscape}
\usepackage{rotating}
\usepackage{graphics}
\usepackage{float}
\usepackage{placeins}
\usepackage{booktabs}
\usepackage{epstopdf,epsfig}
\usepackage{threeparttable}
\usepackage{soul}
\usepackage{enumitem}
\usepackage{tikz}

\usepackage[capitalize]{cleveref}

\floatname{algorithm}{Algorithm}
\biboptions{numbers,sort&compress} 
\journal{XXX}

\newcounter{bla}

\setstcolor{purple}

\begin{document}

\begin{frontmatter}

\title{
Surrogate-assisted airfoil optimization in rarefied gas flows
}

\author{Xiaoda Li} %
\author{Ruifeng Yuan} %
\author{Yanbing Zhang} %
\author{Lei Wu\corref{mycorrespondingauthor}}
  \cortext[mycorrespondingauthor]{Corresponding authors:  \ead{wul@sustech.edu.cn}}
  
\address{Department of Mechanics and Aerospace Engineering, Southern University of Science and Technology, Shenzhen 518055, China}

\begin{abstract}
With growing interest in space exploration, optimized airfoil design has become increasingly important. However, airfoil design in rarefied gas flows remains underexplored because solving the Boltzmann equation formulated in a six-dimensional phase space is time-consuming. To address this problem, a solver-in-the-loop Bayesian optimization framework for symmetric, thickness-only airfoils is developed. First, airfoils are parameterized using a class–shape transformation that enforce geometric admissibility. Second, a Gaussian-process expected-improvement surrogate is coupled in batches to a fast-converging, asymptotic-preserving Boltzmann solver for sample-efficient exploration. Drag-minimizing airfoils are identified in a wide range of gas rarefaction. It is found that, at Mach numbers $M_\infty=2$ and 4, the streamwise force increases with the gas rarefaction and shifts from pressure-dominated to shear-dominated drag, while optimization reduces drag at all conditions. The benefit of optimization peaks in the weakly rarefied regime---about $30\%$ at $M_\infty=2$ and $40\sim50\%$ at $M_\infty=4$---and falls to a few percent in transition and free-molecular flow regimes. Drag decomposition shows that these gains come mainly from reduced pressure drag, with viscous drag almost unchanged. The optimal airfoils form a coherent rarefaction-aware family: they retain a smooth, single-peaked thickness profile, are aft-loaded at low gas rarefaction, and exhibit a forward shift of maximum thickness and thickness area toward mid-chord as gas rarefaction increases. These trends provide a physically interpretable map that narrows the design space.
\end{abstract}


\end{frontmatter}

\section{Introduction}\label{sec:intro}

Shape optimization plays a crucial role in space exploration and high-speed aerospace applications, where vehicle performance, thermal protection, and fuel efficiency are strongly influenced by aerodynamic design. In many regimes encountered during high-altitude flight and atmospheric reentry, the Knudsen number (Kn, defined as the ratio of molecular mean free path to characteristic flow length) becomes appreciable, and the Navier–Stokes (NS) equations break down~\citep{struchtrup2005macroscopic}. 
In these conditions, a kinetic description based on the Boltzmann equation or its model formulations is required to capture the rarefied gas dynamics, which is quite different to continuum gas dynamics. For example, numerical studies show that increasing \(\mathrm{Kn}\) systematically alters lift-to-drag ratio and flow structure~\citep{pekardan2018rarefaction,lofthouse2007effects}. 
However, solving the Boltzmann equation is computationally challenging due to its high dimensionality in phase space, which severely limits the feasibility of routine simulations for design and optimization. As a result, despite its critical importance for the design of spacecraft and hypersonic vehicles~\citep{anderson2006hypersonic,eyi2019shape}, shape optimization in rarefied gas flows remains less developed than in continuum aerodynamics.

Shape optimization approaches can be grouped into two lines: gradient-based local search that relies on sensitivities, and surrogate-model strategies that use a limited number of high-fidelity samples. In the first line, an adjoint system for the chosen flow model is derived and the resulting gradients are used to update high-dimensional design variables efficiently. For continuum flows, gradient-based optimization based on the Navier–Stokes equations has become a standard tool for aerodynamic shape design, where adjoint methods provide gradients of the objective function at a computational cost that is independent of the number of design variables, enabling efficient optimization of wings and airfoils with many design parameters~\citep{jameson1998control,giles2000adjoint,mohammadi2009aso}. For rarefied gas flows, adjoint topology-optimization formulations have recently been constructed for Boltzmann kinetic equations and applied to rarefied microdevices and channels~\citep{sato2019topology,guan2023topology,caflisch2021adjointdsmc,yuan2024design}. These kinetic adjoint approaches, however, require elaborate derivations and intrusive code changes, since boundary conditions and multiscale couplings are algebraically complex and substantial changes in operating conditions or kinetic models often trigger further implementation effort.

In the second line, the mapping from geometry to objective is treated as a black-box function and a surrogate is built on a finite set of training samples to guide subsequent evaluations. Common choices include radial-basis-function interpolants, Gaussian-process models, and neural-network regressors, combined with acquisition rules such as expected improvement, probability of improvement, and upper confidence bounds or related local--global trade-off mechanisms~\citep{rasmussen2006gpml,jones1998ego,rios2013dfo,gardner2014cbo,gonzalez2016localpenal}. Compared with gradient-based methods, these strategies do not require analytic sensitivities, are more robust to non-smooth or noisy objectives, and can be wrapped around different flow solvers without modifying their internal implementation. Their performance, however, is constrained by the number and placement of expensive kinetic evaluations: when each sample demands a high-fidelity solver of the Boltzmann equation~\citep{bird1994dsmc,ivanov1998dsmc,xu2015direct} and the design space has moderate dimension, only a small number of evaluations is affordable and standard surrogate schemes struggle to reconcile global search with local convergence. In addition, if the shape parameterization is not properly regularized, the search tends to generate highly oscillatory airfoils that are difficult to mesh or manufacture~\citep{kulfan2008cst,hicks1978wing}. These limitations naturally raise the question: can the shape space, surrogate model, and sampling strategy be designed in a coordinated way so that, without sacrificing the fidelity of kinetic evaluations and under a limited evaluation budget, the optimal airfoil shapes corresponding to a broad range of Knudsen numbers can be identified?

To address this question, we develop a Bayesian optimization framework for symmetric, thickness-only airfoils in rarefied supersonic flow. We parameterize airfoils using a class–shape transformation (CST) with eight design variables and restrict them to a geometrically admissible set that enforces symmetry, smoothness, and basic thickness and area constraints~\citep{kulfan2008cst,hicks1978wing,samareh2001survey,lyu2014crm}.  Over this regularized shape space, we construct a Gaussian-process surrogate with batched expected-improvement acquisition and incorporate simple feasibility and locality controls~\citep{rasmussen2006gpml,jones1998ego,rios2013dfo,gardner2014cbo,gonzalez2016localpenal}. We couple the surrogate in a solver-in-the-loop framework to an efficient kinetic Boltzmann solver~\citep{su2020gsis,zhu2021gsis,Zhang2023_GSIS}, which provides variance-free evaluations of the dimensionless streamwise force. This approach enables sample-efficient exploration of rarefied airfoil designs and allows us to identify drag-minimizing airfoils from slip to transition and toward the free-molecular flow regime.

The remainder of the paper is organized as follows: Section~\ref{sec:problem} formulates the optimization problem, introduces the CST-based airfoil parameterization and geometric admissible set, and summarizes the mesh generation and kinetic-solver setup. Section~\ref{sec:so} presents the surrogate-based optimization methodology. Section~\ref{sec:results} reports the shape optimization results of the supersonic airfoil over a wide range of Knudsen numbers. Section~\ref{sec:conclusion} concludes and outlines directions for future work.

\section{Problem formulation and numerical setup}\label{sec:problem}

In this section, the airfoil optimization problem is formulated, the CST parameterization and admissible geometries are described, and the mesh generation procedure and multiscale kinetic solver are presented.

\subsection{Objective definition}\label{sec:obj}

Given the airfoil geometry and operating condition (i.e., the Mach number $M_\infty$ and Knudsen number $\mathrm{Kn}$), the wall pressure \(p_w\) and viscous shear-stress tensor \(\boldsymbol{\tau}_w\) on all wall faces can be obtained by solving the Boltzmann kinetic equation. From these fields, the dimensionless streamwise force per unit span, denoted by \(D\), is then constructed and used as the objective in the subsequent shape optimization.

The total wall stress $\boldsymbol{\sigma}_w $ and the traction vector $\mathbf{t}$ on the wall are
\begin{equation}\label{eq:wall-total-stress}
  \boldsymbol{\sigma}_w = \,p_w\,\mathbf{I} + \boldsymbol{\tau}_w, 
  \quad
  \mathbf{t} =\boldsymbol{\sigma}_w \cdot \mathbf{n},
\end{equation}
where \(\mathbf{n}\) is the outward unit normal pointing into the solid. Using the free-stream direction \(\hat{\mathbf{e}}_x=(1,0)\) as the projection axis, the non-dimensional streamwise force per unit span is defined as
\begin{equation}\label{eq:D-continuum}
  D = -\,\frac{1}{p_\infty\,c}\int_{\partial\Omega_w}
 \mathbf{t}\!\cdot\!\hat{\mathbf{e}}_x \,\mathrm{d}s,
\end{equation}
where \(p_\infty=\tfrac{1}{2}\rho_\infty U_\infty^2\) is the free-stream dynamic pressure and \(c\) is the chord length. The minus sign enforces an upwind-positive convention so that \(D>0\) corresponds to drag along the incoming flow.

For non-dimensional geometries we take \(c=1\).  We are going to minimize the drag force at the given set $(M_\infty, \mathrm{Kn})$, when the area of airfoil is no larger than that of the canonical NACA0012 airfoil.

\subsection{CST-based airfoil parameterization and geometric admissibility}\label{sec:cst}

To minimize the objective \(D\) over a tractable design space, the airfoil geometry is represented using the CST formulation. At the same time, the design space is simplified by restricting attention to symmetric, thickness-only airfoils, so that the CST formulation can be simplified accordingly.

Let \(x\in[0,1]\) denote the chordwise coordinate from the leading edge (\(x=0\)) to the trailing edge (\(x=1\)). The upper and lower surfaces of the airfoil can be described as
\begin{equation}
  y_u(x)=\frac{1}{2}\,t(x), 
  \quad 
  y_\ell(x)= -\frac{1}{2}\,t(x),
  \label{eq:surfaces-symm}
\end{equation}
where 
\begin{equation}
  t(x)= 2 x^{N_1}(1-x)^{N_2} \sum_{k=0}^{5}A_k\,B_k^{(5)}(x),
  \label{eq:thickness-symm}
\end{equation}
with $B_k^{(m)}(x)=\binom{m}{k}\,x^{k}(1-x)^{m-k}$ being the Bernstein polynomials and \(\{A_k\}_{k=0}^{5}\)  the Bernstein coefficients. Here, $N_1$ and $N_2$ are positive, and the sharpness or bluntness of the leading and trailing edges is controlled by \(N_1\) and \(N_2\), respectively. 
Each candidate section is therefore fully determined by the six coefficients \(\{A_k\}\) and the two class-function exponents \((N_1,N_2)\), which form the design vector used throughout the optimization:
\begin{equation}
    \boldsymbol{\theta}
=\bigl(A_0,A_1,A_2,A_3,A_4,A_5,N_1,N_2\bigr)\in\mathbb{R}^8.
\end{equation}

To keep the designs within a physically reasonable region, the design variables are first restricted to a simple box in parameter space. In particular, the thickness coefficients \(A_0,\ldots,A_5\) are required to lie in the range of \([0.02,\,0.22]\), and the class-function exponents \(N_1\) and \(N_2\) are confined to the range of \([0.01,\,2.00]\). These bounds are chosen empirically so that the resulting sections bracket the baseline NACA0012 thickness while
excluding needle-like geometries and overly blunt leading or trailing edges. These bounds serve as the global parameter ranges for all optimization runs.

To exclude pathological shapes before meshing and flow simulation, the design vector is further restricted to a geometrically admissible set. All checks are performed on a fixed chordwise grid \(\{x_i\}_{i=0}^{n}\subset[0,1]\).
First, non-intersection of the upper and lower surfaces can be enforced naturally by imposing the constraint $A_k>0$.
Second, the thickness is required to be unimodal. Denoting the maximum thickness and the peak position by $x_{t,\max}={\arg\max}\ t(x)$ and $t_{\max}=t(x_{t,\max})$, respectively,  the following constraints are applied:
\begin{equation}
\left\{ 
\begin{aligned}
     \frac{dt}{dx}>0,  ~\text{for}~ x<x_{t,\max}, \\
     \frac{dt}{dx}<0,  ~\text{for}~ x>x_{t,\max},
\end{aligned}
\right.
  \label{eq:unimodal}
\end{equation}
Third, to limit excessive curvature and small-scale waviness, the curvature of both surfaces is constrained to change sign at most twice along the chord.
Finally, global thickness measures are constrained to avoid needle-like or excessively thick sections. The sectional area $\int_{0}^{1} t(x) dx$ is required to exceed a prescribed minimum area (taken as the area of the canonical NACA0012 in the present non-dimensionalization), and the peak thickness and its location satisfy
\begin{equation}
  \frac{t_{\max}}{c} < 0.25,\quad \frac{x_{t,\max}}{c} \in [0.2,\,0.8].
  \label{eq:tmax-bounds}
\end{equation}

The geometrically admissible set \(\mathcal{A}\) thus consists
of all design vectors \(\boldsymbol{\theta}\) that lie within the parameter ranges and for which the corresponding
surfaces~\eqref{eq:surfaces-symm} satisfy above constraints. This construction removes nonphysical or numerically fragile shapes while retaining sufficient flexibility for meaningful aerodynamic variations.

\subsection{Mesh generation}\label{sec:mesh}

Since the CST parameters are allowed to vary over a broad range, the generated airfoils can become extremely sharp at the leading and trailing edges, which tends to generate nearly degenerate grid cells and, in some cases, even causes failure of the high-fidelity flow solver. Besides, within our sample-intensive optimization workflow, where hundreds of candidate designs must be evaluated one by one, manually adjusting the mesh for each case is infeasible in both time and labor. Therefore, a scripted workflow is designed in the commercial mesh generator Pointwise that automatically identifies the endpoint geometry of each airfoil and selects the corresponding meshing procedure.

\begin{figure}[t!]
  \centering
  \setlength{\tabcolsep}{6pt}
  \renewcommand{\arraystretch}{1.0}
  \begin{tabular}{cc}
    \includegraphics[width=0.45\linewidth]{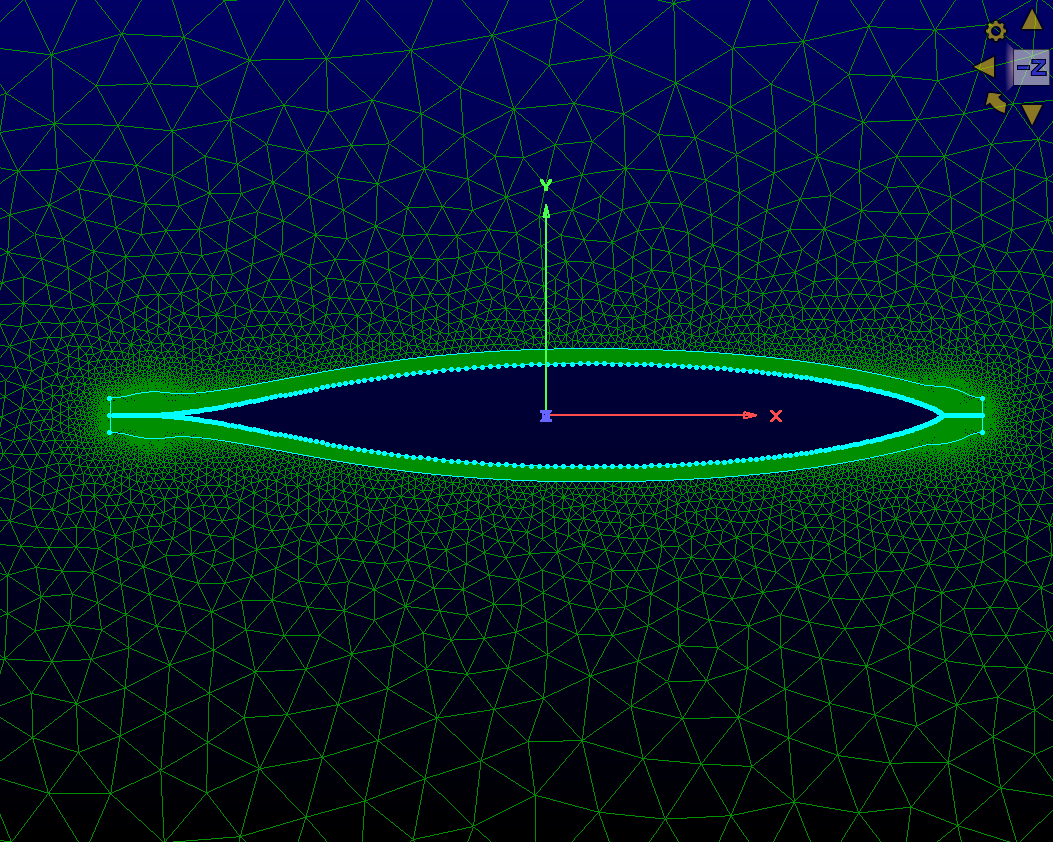} &
    \includegraphics[width=0.45\linewidth]{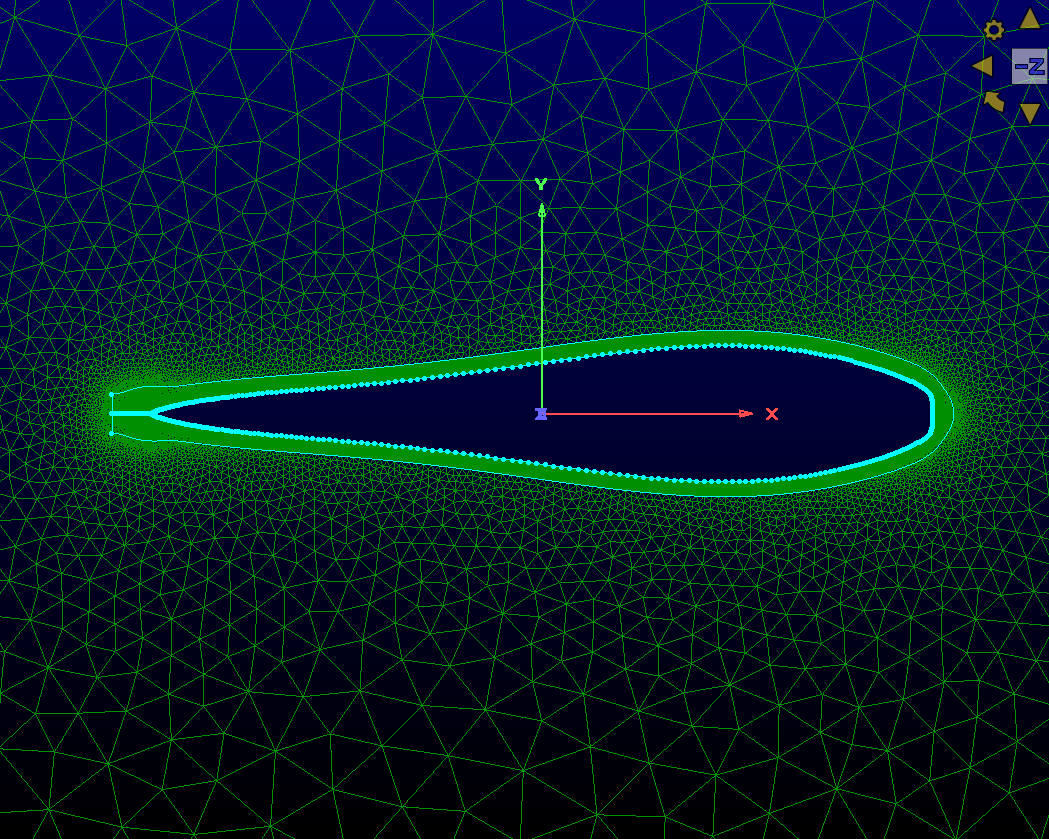} \\
    \small (a) sharp leading edge, sharp trailing edge &
    \small (b) sharp leading edge, blunt trailing edge \\
    \includegraphics[width=0.45\linewidth]{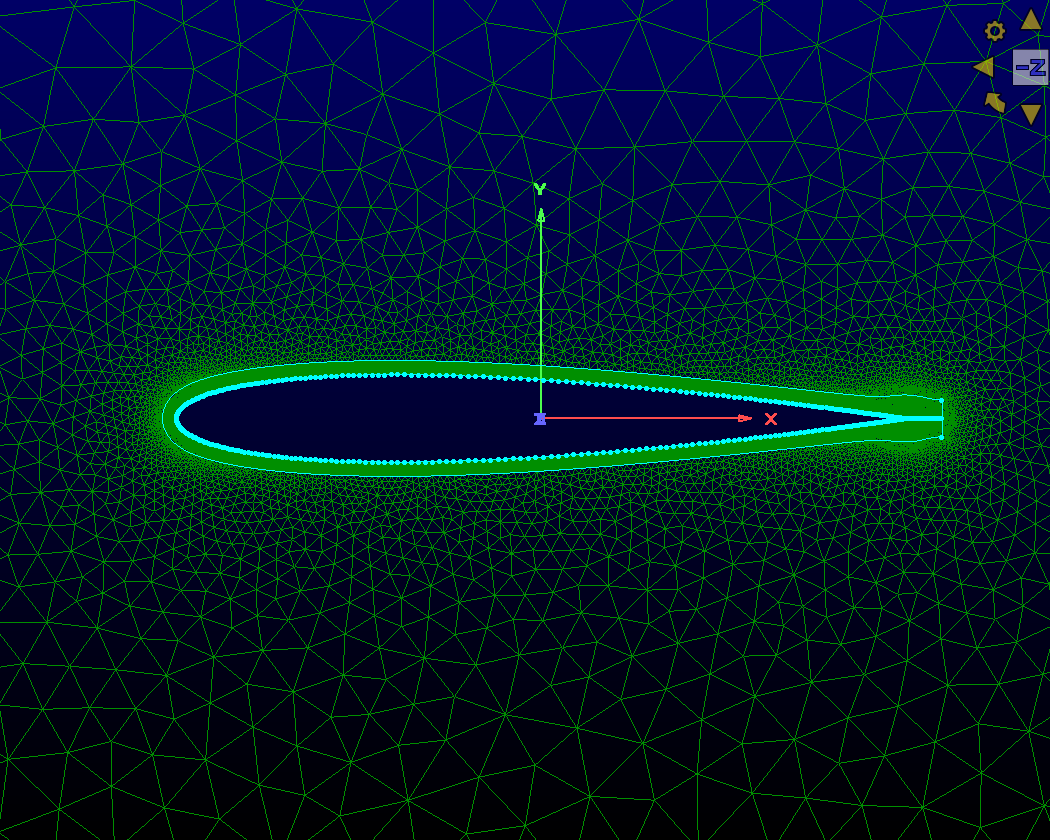} &
    \includegraphics[width=0.45\linewidth]{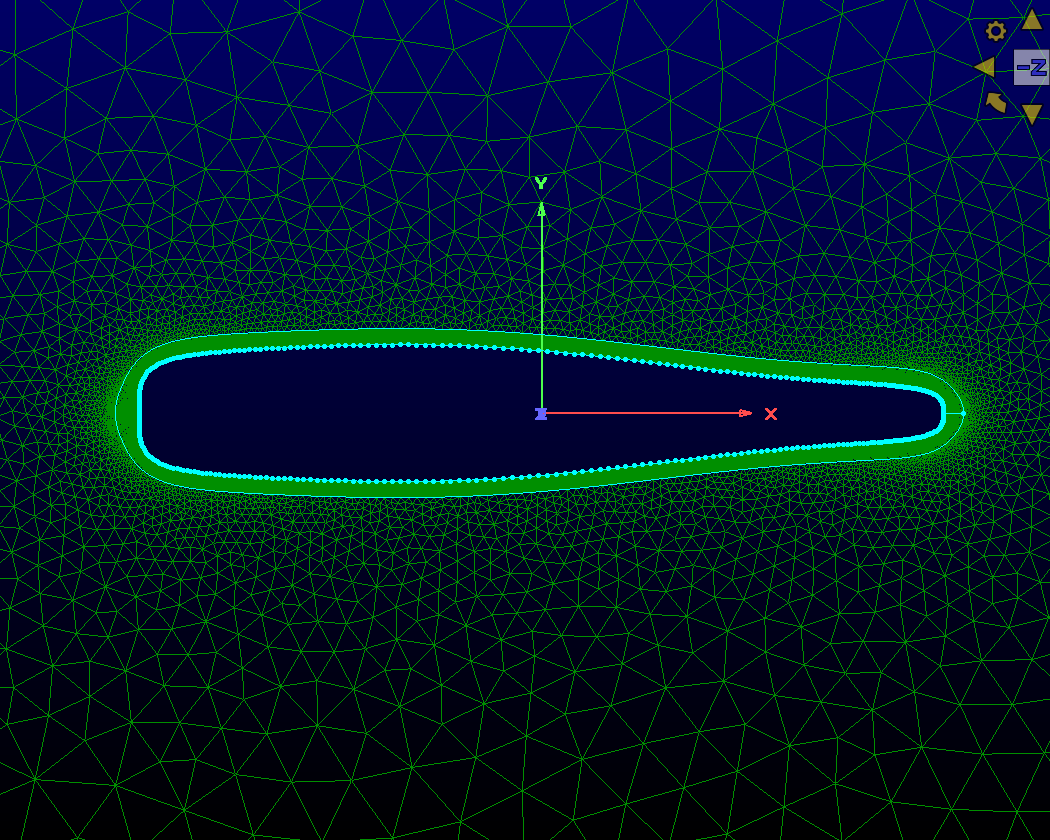} \\
    \small (c) blunt leading edge, sharp trailing edge &
    \small (d) blunt leading edge, blunt trailing edge
  \end{tabular}
  \caption{Representative meshes for the four endpoint topologies.}
  \label{fig:mesh-topos}
\end{figure}

The endpoint geometry is inferred from the CST exponents $N_1$ and $N_2$: if $N_1 ~(\text{or}~ N_2)>0.65$ the leading (or trailing) edge is classified as sharp and otherwise as blunt. Based on this classification, one of four scripts is selected:
\texttt{ss} (sharp leading edge, sharp trailing edge), \texttt{sb} (sharp leading edge, blunt trailing edge),
\texttt{bs} (blunt leading edge, sharp trailing edge), and \texttt{bb} (blunt leading edge, blunt trailing edge).

These four scripts differ in two aspects. On the one hand, for any sharp endpoint, a short tangential cap is added to close the upper and lower surfaces and to provide a well-shaped edge for extrusion; for a blunt endpoint, the CST-generated finite-thickness nose or tail is used directly as the wall boundary. On the other hand, for \texttt{bb}, \texttt{bs} and \texttt{sb} cases, the upper and lower surfaces (including any nose cap) form a single closed boundary that is extruded as one ring. For sharp–sharp (\texttt{ss}) cases, the two sides are instead extruded as separate bands to reduce distortion near sharp trailing edges.

The upper and lower surfaces are reparameterized by arc length and sampled with 201 nodes per side, with endpoint spacings of \(0.0008\,c\) near the nose and tail. This yields nearly uniform sampling and a stable inner curve for the body-fitted strip. The same normal-extrusion parameters are used in all cases: first-layer thickness \(5\times 10^{-4}\,c\), growth factor \(1.04\), and 25 layers, followed by a light smoothing. During this step, the innermost ring is kept fixed and serves as a geometric reference, so that edge ordering, normals, and segment lengths remain consistent with the drag integral in Section~\ref{sec:obj}.
Typical meshes are shown in Fig.~\ref{fig:mesh-topos}.

The far field is a circle of radius \(10\,c\) centered at the airfoil centroid and discretized with 40 nodes. Together with the outer edge of the body-fitted strip, it forms an annular region that is filled with an unstructured mesh. Because the airfoil shape varies between cases, the total number of cells also changes; nevertheless, each grid file contains approximately 20,000 cells. 
In the grid files, the airfoil surface and outer circle are tagged as wall and far field boundaries, where the diffuse and Maxwellian equilibrium boundary conditions are applied, respectively \cite{Zhang2023_GSIS}.

\subsection{Deterministic kinetic solver}\label{sec:solver-gsis}

While every geometrically admissible design vector \(\boldsymbol{\theta}\in\mathcal{A}\) in Section~\ref{sec:cst} is mapped automatically to a CFD-ready mesh, a high-fidelity flow solver is needed to calculate the rarefied gas dynamics. When the Knudsen number becomes appreciable, the NS equations become inaccurate and the gas flow has to be described at the kinetic level.


Consider a polyatomic gas with $d$ internal degrees of freedom.
Two reduced velocity distribution functions, $f_0(\bm{x},\bm{v})$ and $f_1(\bm{x},\bm{v})$, are introduced to describe the translational and internal states of the gas, respectively, where $\bm{x}=(x,y,z)$ is the spatial coordinate and $\bm{v}=(v_x,v_y,v_z)$ is the molecular velocity. 
Denoting the density by $\rho$, the flow velocity by $\bm{u}$, the traceless stress tensor by $\boldsymbol{\sigma}$, the translational and internal temperatures by $T_t$ and $T_r$, and the corresponding heat fluxes by $\bm{q}_t$ and $\bm{q}_r$, one has
\begin{equation}
  \begin{aligned}
    \bigl(\rho,\,\rho\bm{u},\,\boldsymbol{\sigma},\,\tfrac{3}{2}\rho R T_t,\,\bm{q}_t\bigr)
      &= \int_{\mathbb{R}^3} \bigl(1,\,\bm{v},\,\bm{c}\bm{c}-\tfrac{c^2}{3}\boldsymbol{I},\,\tfrac{c^2}{2},\,\tfrac{c^2}{2}\bm{c}\bigr)\, f_0\,\mathrm{d}\bm{v},\\
    \bigl(\tfrac{d}{2}\rho R T_r,\,\bm{q}_r\bigr)
      &= \int_{\mathbb{R}^3} (1,\,\bm{c})\, f_1\,\mathrm{d}\bm{v},
  \end{aligned}
  \label{eq:macro-moments}
\end{equation}
where $\bm{c}=\bm{v}-\bm{u}$ is the peculiar velocity, $\boldsymbol{I}$ is the $3\times 3$ identity tensor, and $R$ is the gas constant. The translational pressure is $p_t=\rho R T_t$, while the total temperature
\(
  T = (3T_t + d T_r)/(3+d)
\)
defines the equilibrium temperature between translational and internal modes, with total pressure $p=\rho R T$.

The evolution of the reduced VDFs is governed by the modified Rykov model~\citep{rykov1975model,wu2015kinetic}, which in the present steady-state problems takes the form
\begin{equation}
  \bm{v}\cdot\nabla_{\bm{x}} f_i
=
  \frac{g_{i,t}-f_i}{\tau}
  + \frac{g_{i,r}-g_{i,t}}{Z_r\,\tau},
  \quad i=0,1,
  \label{eq:boltzmann-gsis}
\end{equation}
where $g_{i,t}$ and $g_{i,r}$ are reference (translational and internal) equilibrium distributions, $Z_r = 2.667$ is the internal collision number that controls the relaxation of internal energy toward translational equilibrium, and $\tau$ is the mean collision time of gas molecules. The specific forms of $g_{0,t}$, $g_{0,r}$, $g_{1,t}$ and $g_{1,r}$ are chosen such that the model reproduces the correct shear viscosity and thermal conductivity for diatomic gases, see Refs.~\cite{wu2015kinetic,li2021uncertainty} for details.

The collision time is related to the local shear viscosity and pressure through
\begin{equation}
  \tau = \frac{\mu(T_t)}{p_t},
  \quad
  \mu(T_t) = \mu(T_0)\left(\frac{T_t}{T_0}\right)^{\omega},
  \label{eq:tau-mu}
\end{equation}
where $\mu(T_t)$ is the gas viscosity, $T_0$ is a reference temperature, and $\omega$ is the viscosity index associated with the intermolecular potential; in this paper, we choose $\omega=0.75$ for nitrogen gas, with $d=2$. The Knudsen number is defined at the reference pressure $p_0$ and temperature $T_0$
as 
\begin{equation}
  \mathrm{Kn}
=
  \frac{\mu(T_0)}{p_0 L}\,
  \sqrt{\frac{\pi R T_0}{2}}.
  \label{eq:kn-viscosity}
\end{equation}


The kinetic model above is formulated with three translational degrees of freedom, but in the numerical implementation we restrict attention to planar flows and assume that the distribution function is independent of the out-of-plane velocity component. As a result, the three-dimensional velocity integrals reduce to integrals over the in-plane components $(v_x,v_y)$ and are evaluated on a truncated two-dimensional velocity domain
\(
  (v_x,v_y)\in[-V_{x,\max},V_{x,\max}]
               \times[-V_{y,\max},V_{y,\max}],
\)
and discretized by a uniform Cartesian grid with
$N_{v_x}\times N_{v_y}$ nodes. Specifically, we use
\begin{equation}
    \begin{aligned}
        (V_{x,\max},V_{y,\max},N_{v_x},N_{v_y})&
  = (10,\,8,\,50,\,40)
  \quad\text{for } M_\infty=2, \\
   (V_{x,\max},V_{y,\max},N_{v_x},N_{v_y})&
  = (14,\,12,\,70,\,60)
  \quad\text{for } M_\infty=4.
    \end{aligned}
\end{equation}
For each Mach number, this velocity grid is kept fixed over all Knudsen numbers considered, and velocity moments such as density, stress, and heat flux are approximated by simple quadrature over the corresponding discrete velocity set~\cite{Zhang2023_GSIS}.

Since conventional discrete-velocity methods tend to converge slowly when the flow approaches the near-continuum regime, the general synthetic iterative scheme (GSIS)~\citep{su2020gsis,zhu2021gsis,Zhang2023_GSIS} is adopted to accelerate convergence. GSIS is built on a discrete-velocity, finite-volume discretization of Eq.~\eqref{eq:boltzmann-gsis}, but augments it with a synthetic macroscopic system rigorously derived from the kinetic equation. That is, at each iteration, macroscopic moments $\bm{W} = (\rho,\bm{u},T)$, representing the conservative densities of mass, momentum, and total energy, are first evaluated from the current kinetic solution. An auxiliary macroscopic system is then solved to obtain an updated field $\bm{W}$ that is asymptotically consistent with the NS equations in the small-$\mathrm{Kn}$ limit. This updated macroscopic field is fed back to precondition and constrain the kinetic update, yielding new velocity distribution functions $f_0$ and $f_1$. The process is repeated until both $f$ and $\bm{W}$ converge. This synthetic macro–meso coupling enables fast-converging, asymptotic-preserving simulations of rarefied gas flows, typically reaching steady state within a few dozen iterations while allowing spatial cell sizes much larger than the molecular mean free path.
As a consequence, a steady GSIS solve for a single airfoil at one operating condition $(M_\infty,\mathrm{Kn})$ typically requires about $2 \sim 3$ minutes of wall-clock time for $M_\infty=2$ when run on 60 cores~\cite{Zhang2023_GSIS}.

An isothermal diffuse-reflection boundary condition is imposed at the solid wall \cite{zhu2021gsis}, where the wall temperature equals that of the incoming flow at far field. To evaluate the drag objective, the stresses acting on the airfoil surface are obtained from the kinetic solution at the solid wall. Since GSIS is deterministic and variance-free, quantities such as drag remain stable across iterations and designs, which is crucial for robust comparison in optimization.
\section{Surrogate Optimization}\label{sec:so}

\begin{figure}[t!]
  \centering
  \includegraphics[width=0.65\linewidth]{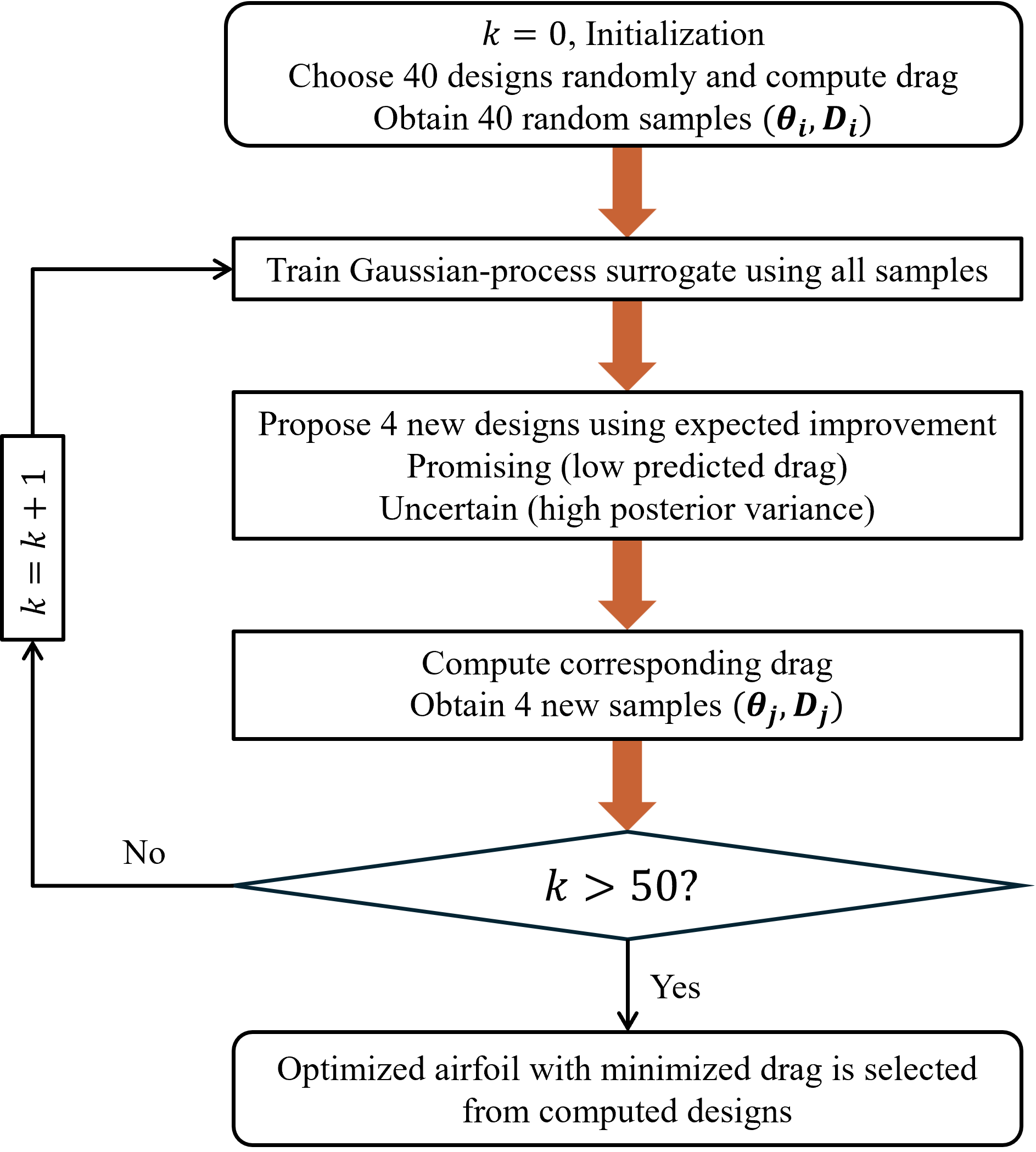}
  \caption{Surrogate-assisted optimization workflow at a fixed operating condition $(M_\infty,\mathrm{Kn})$.}
  \label{fig:opt-workflow}
\end{figure}

The solver-in-the-loop workflow to minimize the drag is shown in Fig.~\ref{fig:opt-workflow}.
The loop is built around a Gaussian-process surrogate with automatic relevance determination (ARD)~\citep{rasmussen2006gpml} in Section~\ref{sec:gp-ard}, which is initially trained on 40 admissible airfoil designs generated by Latin hypercube sampling~\citep{mckay1979lhs} and evaluated by GSIS. In each of 50 subsequent optimization rounds, an expected improvement (EI) acquisition rule~\citep{jones1998ego} is used to score candidate designs, and a batch of four new designs is selected for high-fidelity GSIS evaluation and appended to the training set (Section~\ref{sec:acq}). To improve robustness and accelerate convergence under this limited evaluation budget, the feasibility-aware filters, box-based locality controls, and simple parameter-tuning strategies that modify the candidate generation, scoring, and batch-selection, is introduced in Section~\ref{sec:acq-engineering}. The choices of 40 initial designs and 50 optimization rounds are based on empirical tests that balance surrogate accuracy against overall computational cost.

\subsection{Gaussian-process with ARD}\label{sec:gp-ard}

For a fixed operating condition \((M_\infty,\mathrm{Kn})\), the design vector introduced in Section~\ref{sec:cst} is denoted by \(\boldsymbol{\theta}\in\mathcal{A}\) and is represented here by the raw input vector \(\tilde{\boldsymbol{x}}\in\mathbb{R}^8\), i.e.\ \(\tilde{\boldsymbol{x}} \equiv \boldsymbol{\theta}\). The corresponding target is the steady drag \(\tilde{y}=D(\boldsymbol{\theta};M_\infty,\mathrm{Kn})\).

For the design variables, it is convenient to apply a simple normalization so that all components share a comparable scale and the minimum distance filters and trust region updates can be formulated in a common coordinate system. Let \(\text{lo}_j\) and \(\text{hi}_j\) be the componentwise minima and maxima of the raw inputs over the current training set. Each component is then mapped to a normalized coordinate by \(x_j = (\tilde{x}_j - \text{lo}_j)/(\max(\text{hi}_j - \text{lo}_j, 10^{-6}))\), where a lower bound of \(10^{-6}\) is added to avoid division by very small numbers.

The scalar outputs are treated differently. To improve the numerical conditioning of the
Gaussian-process regression and to match a zero-mean prior, the raw responses \(\tilde{y}\)
are standardized rather than merely rescaled. Specifically, the standardized output is
defined as \(y = (\tilde{y}-\bar{y})/s_y\), where \(\bar{y}\) and \(s_y\) are the sample
mean and standard deviation of the raw outputs. With this treatment, \(y\) has
approximately zero mean and unit variance and is well conditioned for hyperparameter estimation.

On the normalized design space, let \(y(\boldsymbol{x})\) denote the standardized drag
response at input \(\boldsymbol{x}\). A Gaussian-process prior is placed on this scalar
response,
\begin{equation}
  y(\cdot)\sim\mathcal{GP}\bigl(0,\,k_\eta(\cdot,\cdot)\bigr),
  \label{eq:gp-prior}
\end{equation}
where the first argument \(0\) is the prior mean function, chosen to be identically zero
because the standardized outputs have approximately zero mean. The covariance kernel
\(k_\eta\) encodes similarity and smoothness. To allow different sensitivities along
different CST parameters, a Matérn–\(5/2\) kernel with automatic relevance determination
(ARD) is used~\citep{rasmussen2006gpml}:
\begin{equation}
  k_\eta(\boldsymbol{x},\boldsymbol{x}')
  =\sigma^2\!\left(1+\sqrt{5}\,r+\tfrac{5}{3}r^2\right)\exp\bigl(-\sqrt{5}\,r\bigr),
  \label{eq:m52-ard}
\end{equation}
where $r(\boldsymbol{x},\boldsymbol{x}')  =\sqrt{\sum_{j=1}^{8}{(x_j-x'_j)^2}/{\ell_j^2}}$ is the scaled Euclidean distance in the normalized design space.
Here the \(\ell_j\) are ARD length scales that control the sensitivity of the response to
the \(j\)-th CST parameter: smaller \(\ell_j\) permit more rapid variation along that
direction, while larger \(\ell_j\) correspond to weaker dependence. The parameter
\(\sigma^2\) is the process variance that sets the overall magnitude of the
function variations. In the implementation, all length scales are initialized to
\(\ell_j=1\) and the process variance is initialized to \(\sigma^2=1\), with bounds \(\ell_j\in[10^{-3},10^{5}]\) and
\(\sigma^2\in[10^{-4},10^{4}]\). These quantities are treated as hyperparameters of the surrogate model and will be learned and updated from the training data as the optimization proceeds.

Given the training data
\(X=[\boldsymbol{x}_1,\ldots,\boldsymbol{x}_n]^\top\) and
\(\boldsymbol{y}=[y(\boldsymbol{x}_1),\ldots,y(\boldsymbol{x}_n)]^\top\),
the Gaussian-process posterior for the standardized drag at any test point
\(\boldsymbol{x}_\ast\) has mean and variance
\begin{equation}
  \mu(\boldsymbol{x}_\ast)=\boldsymbol{k}_\ast^\top K^{-1}\boldsymbol{y},
  \quad
  s^2(\boldsymbol{x}_\ast)
  =\sigma^2-\boldsymbol{k}_\ast^\top K^{-1}\boldsymbol{k}_\ast,
  \label{eq:gp-posterior}
\end{equation}
where
\(\boldsymbol{k}_\ast=[k_\eta(\boldsymbol{x}_\ast,\boldsymbol{x}_1),\ldots,
k_\eta(\boldsymbol{x}_\ast,\boldsymbol{x}_n)]^\top\)
and \(K\in\mathbb{R}^{n\times n}\) is the kernel matrix collecting the covariances among
the training inputs. To represent small numerical noise and to stabilize the inversion of the covariance
matrix, a diagonal term is added and \(K\) is defined in practice as $K_{ij}=k_\eta(\boldsymbol{x}_i,\boldsymbol{x}_j)+\sigma_n^2\,\delta_{ij},$ where \(\delta_{ij}\) denotes the Kronecker delta and \(\sigma_n^2\) is an observation noise variance. In the implementation, \(\sigma_n^2\) is initialized to \(10^{-6}\) with bounds \(\sigma_n^2\in[10^{-12},10^{-2}]\), and is learned from the data together with the other kernel hyperparameters.

These kernel and noise parameters are collected in a vector of hyperparameters \(\boldsymbol{\eta}=\{\sigma,\{\ell_j\}_{j=1}^{8},\sigma_n\}\), which are learned by maximizing the log marginal likelihood~\citep{jones1998ego,rasmussen2006gpml}:
\begin{equation}
  \log p(\boldsymbol{y}\mid X,\boldsymbol{\eta})
  = -\tfrac12\boldsymbol{y}^\top K^{-1}\boldsymbol{y}
    -\tfrac12\log(\det(K))-\tfrac{n}{2}\log(2\pi).
  \label{eq:gp-logml}
\end{equation}

Once trained, this Gaussian-process surrogate provides a Gaussian predictive distribution for the standardized latent drag \(y(\boldsymbol{x}_\ast)\,\big|\,X,\boldsymbol{y},\boldsymbol{\eta} \sim \mathcal{N}\bigl(\mu(\boldsymbol{x}_\ast),\,s^2(\boldsymbol{x}_\ast)\bigr)\) for any normalized input $\boldsymbol{x}_\ast$, with mean and variance given by Eq.~\eqref{eq:gp-posterior}. The corresponding prediction
for the drag on the original scale is obtained by the inverse standardization
\(\tilde{y}(\boldsymbol{x}_\ast)=\bar{y}+s_y\,y(\boldsymbol{x}_\ast)\). These predictions serve both as a smooth surrogate of the high-fidelity GSIS evaluations and as the source of the mean and uncertainty information used by the acquisition and engineering refinements in Sections~\ref{sec:acq} and~\ref{sec:acq-engineering}.

\subsection{Acquisition and Scoring: Uncertainty-Guided Selection}\label{sec:acq}

To improve surrogate accuracy and guide the search toward low-drag optima, each optimization round proposes four new designs that are predicted by the Gaussian-process model to be both promising (low predicted drag) and uncertain (high predictive variance). In each round, a finite candidate pool \(\mathcal{C}\subset\mathcal{A}\)  is generated, and shapes that are not manufacturable or that tend to produce unstable meshes or solver failures are discarded. The remaining feasible candidates are then scored and ranked using the Gaussian-process predictive distribution, and the top four designs are selected.

Let the current training data be $\mathcal{D}=\{(\boldsymbol{x}_i,y_i)\}_{i=1}^n$ and the best objective value  be $ y^{+} = \min_{1\le i\le n} y_i$. With the Gaussian-process surrogate in Section~\ref{sec:gp-ard}, the predictive response at any candidate $\boldsymbol{x}$ is modeled by a normal random variable $Y(\boldsymbol{x})\sim\mathcal{N}\bigl(\mu(\boldsymbol{x}),s^2(\boldsymbol{x})\bigr)$, with $\mu$ and $s$ being the predictive mean and standard deviation, respectively.
For minimization, the potential gain over the current best is measured by the improvement random variable $I(\boldsymbol{x})=\max\{0,\,y^{+}-Y(\boldsymbol{x})\}$. The corresponding expected improvement (EI) has the closed form~\citep{jones1998ego}:
\begin{equation}
  \mathrm{EI}(\boldsymbol{x}) = \bigl(y^{+}-\mu(\boldsymbol{x})\bigr)\,\Phi\!\bigl(z(\boldsymbol{x})\bigr)
  + s(\boldsymbol{x})\,\phi\!\bigl(z(\boldsymbol{x})\bigr),
  \quad \text{with}~
  z(\boldsymbol{x})=\frac{y^{+}-\mu(\boldsymbol{x})}{s(\boldsymbol{x})},
  \label{eq:ei}
\end{equation}
where \(\Phi\) and \(\phi\) denote the standard normal cumulative distribution function
and probability density function, respectively.

Since the magnitudes of $\mathrm{EI}$, $s$, and $\mu$ may vary across operating conditions and optimization rounds, these quantities are rescaled within each candidate pool $\mathcal{C}$ to yield comparable, dimensionless scores. Specifically, for $\boldsymbol{x}\in\mathcal{C},$ the pool-wise min--max normalization is given by
\begin{equation}
\begin{aligned}
  \widehat{\mathrm{EI}}(\boldsymbol{x}) =&
  \frac{\mathrm{EI}(\boldsymbol{x})-\min_{\boldsymbol{u}\in\mathcal{C}}\mathrm{EI}(\boldsymbol{u})}
       {\max_{\boldsymbol{u}\in\mathcal{C}}\mathrm{EI}(\boldsymbol{u})-\min_{\boldsymbol{u}\in\mathcal{C}}\mathrm{EI}(\boldsymbol{u})+\varepsilon},\\
  \widehat{s}(\boldsymbol{x}) =&
  \frac{s(\boldsymbol{x})-\min_{\boldsymbol{u}\in\mathcal{C}}s(\boldsymbol{u})}
       {\max_{\boldsymbol{u}\in\mathcal{C}}s(\boldsymbol{u})-\min_{\boldsymbol{u}\in\mathcal{C}}s(\boldsymbol{u})+\varepsilon},\\
  \widehat{(-\mu)}(\boldsymbol{x}) =&
  \frac{-\mu(\boldsymbol{x})-\min_{\boldsymbol{u}\in\mathcal{C}}\{-\mu(\boldsymbol{u})\}}
       {\max_{\boldsymbol{u}\in\mathcal{C}}\{-\mu(\boldsymbol{u})\}-\min_{\boldsymbol{u}\in\mathcal{C}}\{-\mu(\boldsymbol{u})\}+\varepsilon},
\end{aligned}
\end{equation}
with a small positive $\varepsilon$ for numerical stability. The three normalized signals $\widehat{\mathrm{EI}}$, $\widehat{s}$, and $\widehat{(-\mu)}$ take values in $[0,1]$ and represent, respectively, improvement potential, uncertainty, and preference for low predicted drag.

For each candidate pool $\mathcal{C}$, the composite score used in the basic acquisition rule is defined as
\begin{equation}
  S_0(\boldsymbol{x}) \;=\; 0.85\,\widehat{\mathrm{EI}}(\boldsymbol{x})
  + 0.15\,\widehat{s}(\boldsymbol{x}),
  \label{eq:S0}
\end{equation}
where the two weights form a simple convex combination that gives EI primary influence while retaining a modest contribution from posterior uncertainty and their precise values are chosen heuristically.

\subsection{Locality Controls for Accelerated Convergence}\label{sec:acq-engineering}

To improve robustness and accelerate convergence under the limited evaluation budget, several additional controls are imposed on the acquisition scores and candidate generation, in the spirit of batch designs that promote diversity and trust region strategies for Bayesian optimization~\citep{gonzalez16localpenal,eriksson19turbo}. Here the iteration progress $t\in[0,1]$ is defined as the fraction of completed optimization rounds, and these additional controls are activated only when $t$ exceeds prescribed thresholds.

Firstly, when $t \ge 0.50$, the score in Eq.~\eqref{eq:S0} is modified to introduce a low-mean bias:
\begin{equation}
  S(\boldsymbol{x}) = 0.4S_0(\boldsymbol{x})
  + 0.6\bigl[\,1-\widehat{\mu}(\boldsymbol{x})\,\bigr].
  \label{eq:S}
\end{equation}
This promotes candidates with lower predicted drag as the search proceeds, while retaining a nonzero contribution from $S_0$ so that designs with high EI and uncertainty can still be explored and selected.

Besides, during this selection described in Section~\ref{sec:acq}, each new design is further required to keep at least a prescribed distance from all designs that have already entered the batch in the normalized space, so that the limited evaluations are not wasted on nearly duplicate designs. To allow later-stage batches to explore the most promising region more densely and thereby
promote convergence, the minimum separation between batch members is gradually reduced as
the optimization progresses. Specifically, we prescribe the progress-dependent minimum distance in the normalized
design space as
\begin{equation}
  \delta(t)=
  \begin{cases}
    0.03, & t<0.30\\[2pt]
    0.02, & 0.30\le t<0.60\\[2pt]
    0.01, & t\ge 0.60
  \end{cases}
  \label{eq:min-dist}
\end{equation}
A candidate with normalized design vector \(\boldsymbol{x}\)
is eligible for inclusion in the batch only if the distance constraint $ \bigl\|\boldsymbol{x}-\boldsymbol{u}_{k}\bigr\|_2 \;\ge\; \delta(t)$ is satisfied for all normalized design vectors \(\boldsymbol{u}_{k}\) of the previously
selected batch members. 

Finally, when evidence for a promising region becomes strong, sampling from the global
box is further restricted to an axis-aligned box around the current best designs.
Specifically, when \(t\ge 0.50\) and at least ten ground-truth evaluations are available,
let \(\mathcal{S}\) denote the set of the ten designs with the smallest objective values.
For each design coordinate \(j\), we define
\begin{equation}
  [\text{lo}_{\mathrm{eff},j},\,\text{hi}_{\mathrm{eff},j}] =
  \Bigl[\min_{\boldsymbol{x}\in\mathcal{S}} x_j,\ \max_{\boldsymbol{x}\in\mathcal{S}} x_j\Bigr],
  \label{eq:topk-box}
\end{equation}
where \(x_j\) denotes the \(j\)th component of a design vector \(\boldsymbol{x}\). 
Thus \(\text{lo}_{\mathrm{eff},j}\) and \(\text{hi}_{\mathrm{eff},j}\) are the
coordinate-wise minima and maxima over the ten best designs and act as effective lower
and upper bounds along coordinate \(j\) for candidate generation in the current round.
The resulting local box is used only for candidate generation, while the Gaussian-process
coordinates and normalization remain unchanged so that the surrogate model is trained and
queried on a consistent global input space. As the optimization progresses and the
membership of \(\mathcal{S}\) evolves, these bounds are recomputed, so that the sampling region is dynamically adapted and gradually contracted around
the region containing the currently best designs, which promotes faster convergence of
the overall optimization.

\section{Numerical results}\label{sec:results}

In this section, a mesh-independence study is first conducted to verify the numerical setup, followed by validation of the surrogate-assisted optimization framework against reference solutions. Then, the framework is applied to identify, at \(M_\infty=2\) and \(M_\infty=4\), drag-minimizing airfoils over a wide range of Knudsen numbers. Finally, the resulting optimal designs are used to characterize their Knudsen- and Mach-number dependence through the behavior of drag and key geometric features. In all test cases, the GSIS solver is run with identical numerical settings, except that the velocity-space domain and discretization are adjusted with the freestream Mach  number, as detailed in Section~\ref{sec:solver-gsis}.

\subsection{Mesh-independence verification}\label{sec:mesh2}

\begin{table}[t!]
  \centering
  \caption{Objective \(D\) for NACA~0012 at
  \(\mathrm{Ma}=2\) under four representative Knudsen numbers. Grids G1/G2/G3 use
  201/301/401 chord-wise nodes on the upper and lower surfaces, 25/35/50 near-wall
  layers with first-layer thicknesses
  \(5\times10^{-4}c/3.5\times10^{-4}c/2.5\times10^{-4}c\) and growth factors
  1.04/1.035/1.03; the far-field radius and sparsity are fixed at \(10c\) and
  40 nodes.}
  \label{tab:mesh-independence}
  \begin{tabular}{lccc}
    \toprule
    \(\mathrm{Kn}\) & G1 (coarse) & G2 (medium) & G3 (fine) \\
    \midrule
    \(0.001\)  & 0.4466 & 0.4427 & 0.4380 \\
    \(0.01\)   & 0.8002 & 0.7936 & 0.7889 \\
    \(0.1\)    & 1.5559 & 1.5509 & 1.5495 \\
    \(1.0\)    & 2.1730 & 2.1720 & 2.1713 \\
    \bottomrule
  \end{tabular}
\end{table}

We first verify mesh independence of the objective \(D\) using NACA~0012 at
\(\mathrm{Ma}=2\) under the four representative Knudsen numbers listed in
Table~\ref{tab:mesh-independence}. 
Numerically, \(D\) decreases monotonically from G1 (coarse) to G3 (fine) for all four
Knudsen numbers, with the coarse-to-fine change remaining below about \(2\%\) and the
medium-to-fine change decreasing from approximately \(1.1\%\) at
\(\mathrm{Kn}=10^{-3}\) to \(0.03\%\) at \(\mathrm{Kn}=1\). 
In view of the small coarse-to-fine differences (at most about \(2\%\)) and the large number of samples required in the optimization studies, the coarse grid G1 is adopted in the remainder of this work.

\subsection{Validation of the optimization framework}\label{sec:result-ma2}

To validate the surrogate-assisted optimization framework, shape optimizations are
performed at \(\mathrm{Kn}=0.01\) and \(\mathrm{Kn}=0.50\) with a freestream Mach number
of \(M_\infty=2\), and the resulting optimal airfoils are compared with the adjoint-based
designs~\cite{yuan2025adjoint}. To avoid inconsistencies in
objective definition, nondimensionalization, and boundary modeling, we do not compare
absolute drag values here, but focus on the changes in geometry and flow field from the
initial to the optimized airfoils. 

\begin{table}[h]
  \centering
  \caption{ Summary of surrogate-assisted airfoil optimizations at \( \mathrm{Ma}=2 \). Each airfoil mesh contains some \(2\times10^4\) cells. The “Time” column reports the total wall-clock time for all \(240\) samples per case on 60 cores, corresponding to roughly 2~min per steady GSIS solve~\cite{Zhang2023_GSIS} at \(\mathrm{Kn}=0.01\) and 3~min at \(\mathrm{Kn}=0.50\).  }
  \label{tab:ma2_summary}
  \small
  \begin{tabular}{cccccc}
    \toprule
    Kn & $D$ (initial) & $D$ (optimized) & Reduction  & Time \\
    \midrule
    0.01 & 0.8054 & 0.7151 & 11.2\%  & 8 hours \\
    0.50 & 2.0670 & 1.9706 & 4.7\%  & 12 hours \\
    \bottomrule
  \end{tabular}
  \normalsize
\end{table}

As summarized in Table~\ref{tab:ma2_summary}, the optimized airfoils reduce the drag by about \(11\%\) at \(\mathrm{Kn}=0.01\) and \(5\%\) at \(\mathrm{Kn}=0.50\). In both cases, candidates very close to the final optimum already appear after about
150 samples. The eventual choice of 240 samples (see Section~\ref{sec:so}) is intended
to improve statistical robustness, so that the ten lowest-drag airfoils in the computed
set are nearly indistinguishable in shape.

\begin{figure}[t!]
  \centering
  \includegraphics[width=.48\linewidth]{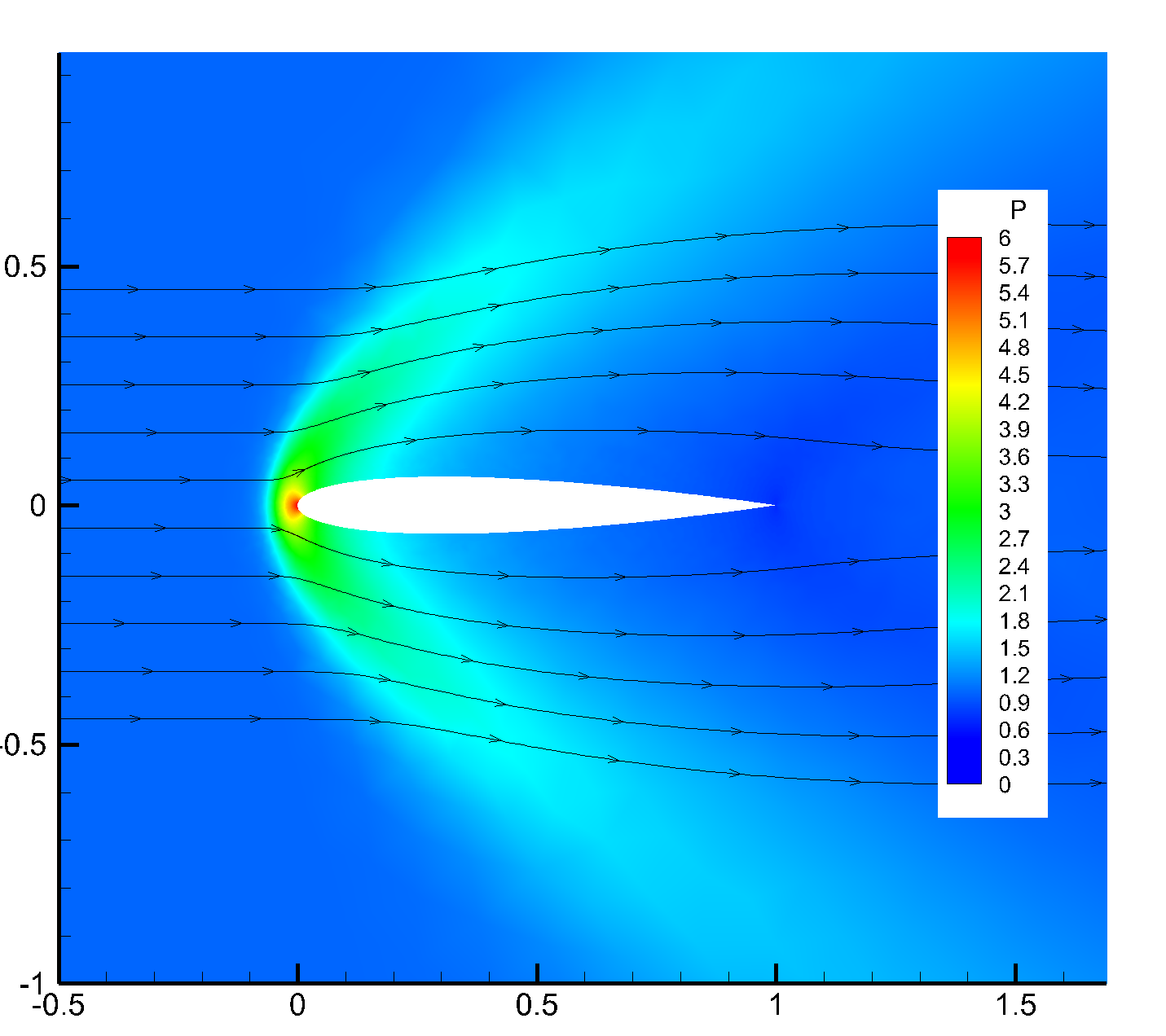}
  \hspace{0.001\linewidth}
  \includegraphics[width=.48\linewidth]{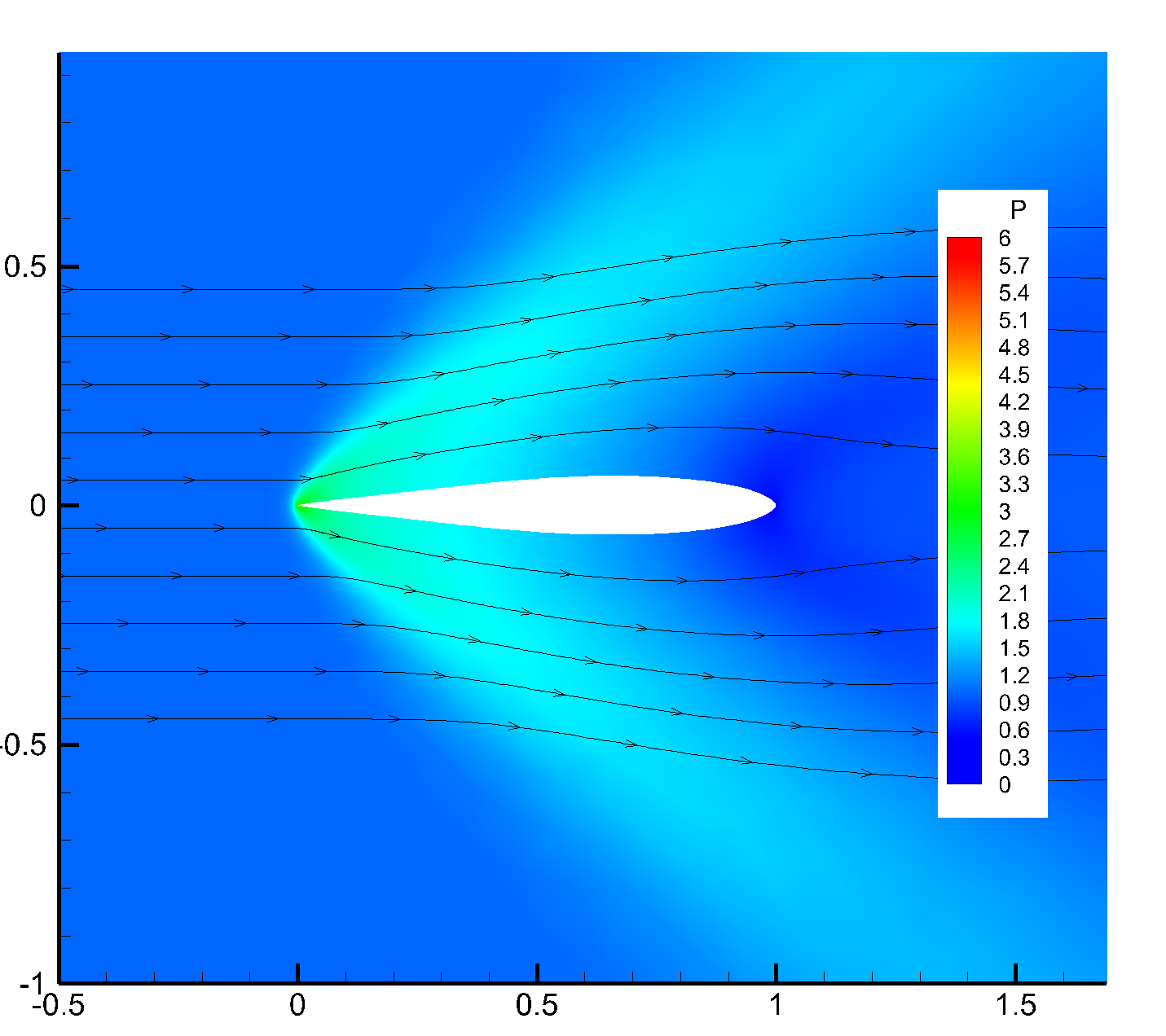}\\[2pt]
  \includegraphics[width=.48\linewidth]{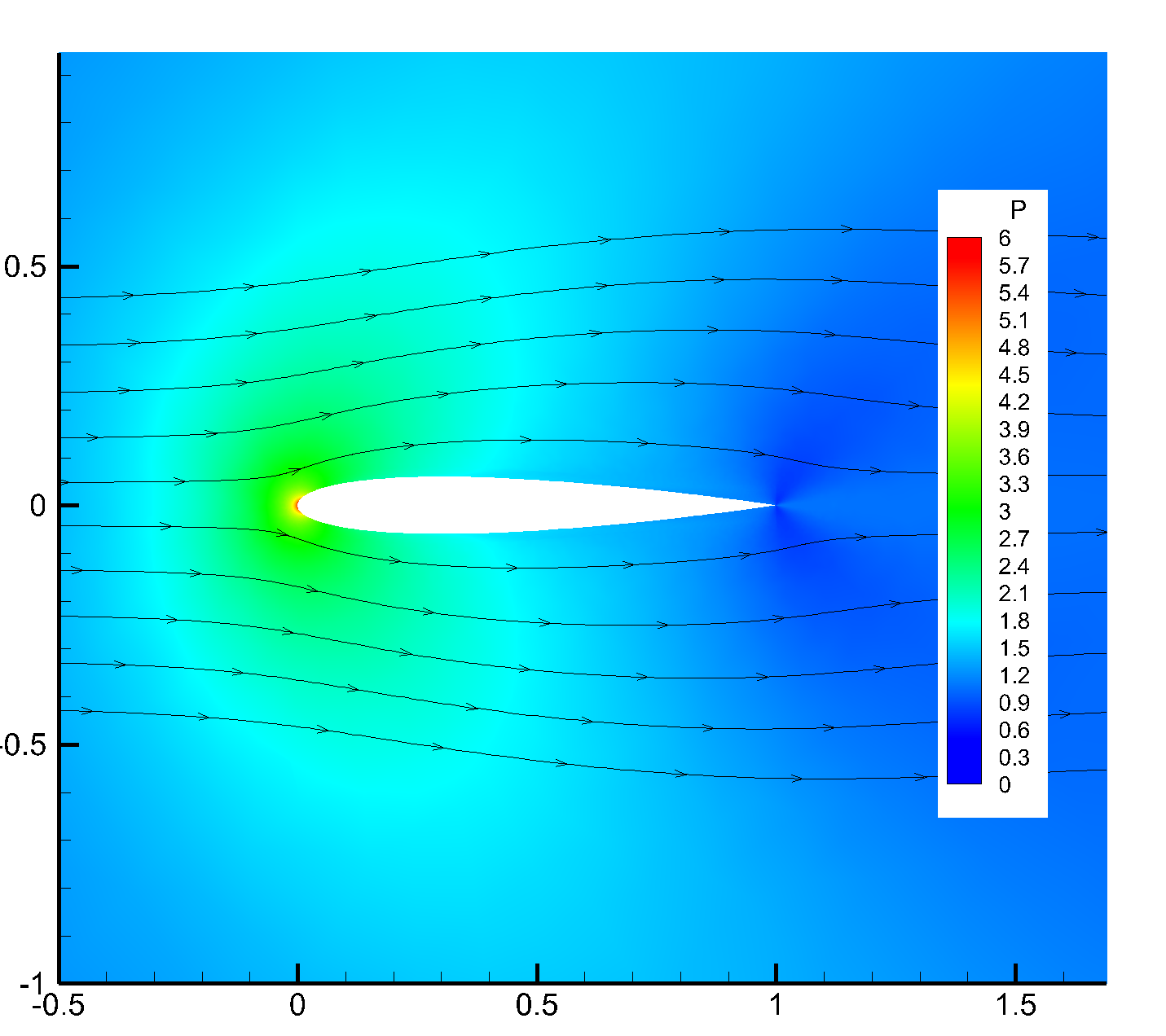}
  \hspace{0.001\linewidth}
  \includegraphics[width=.48\linewidth]{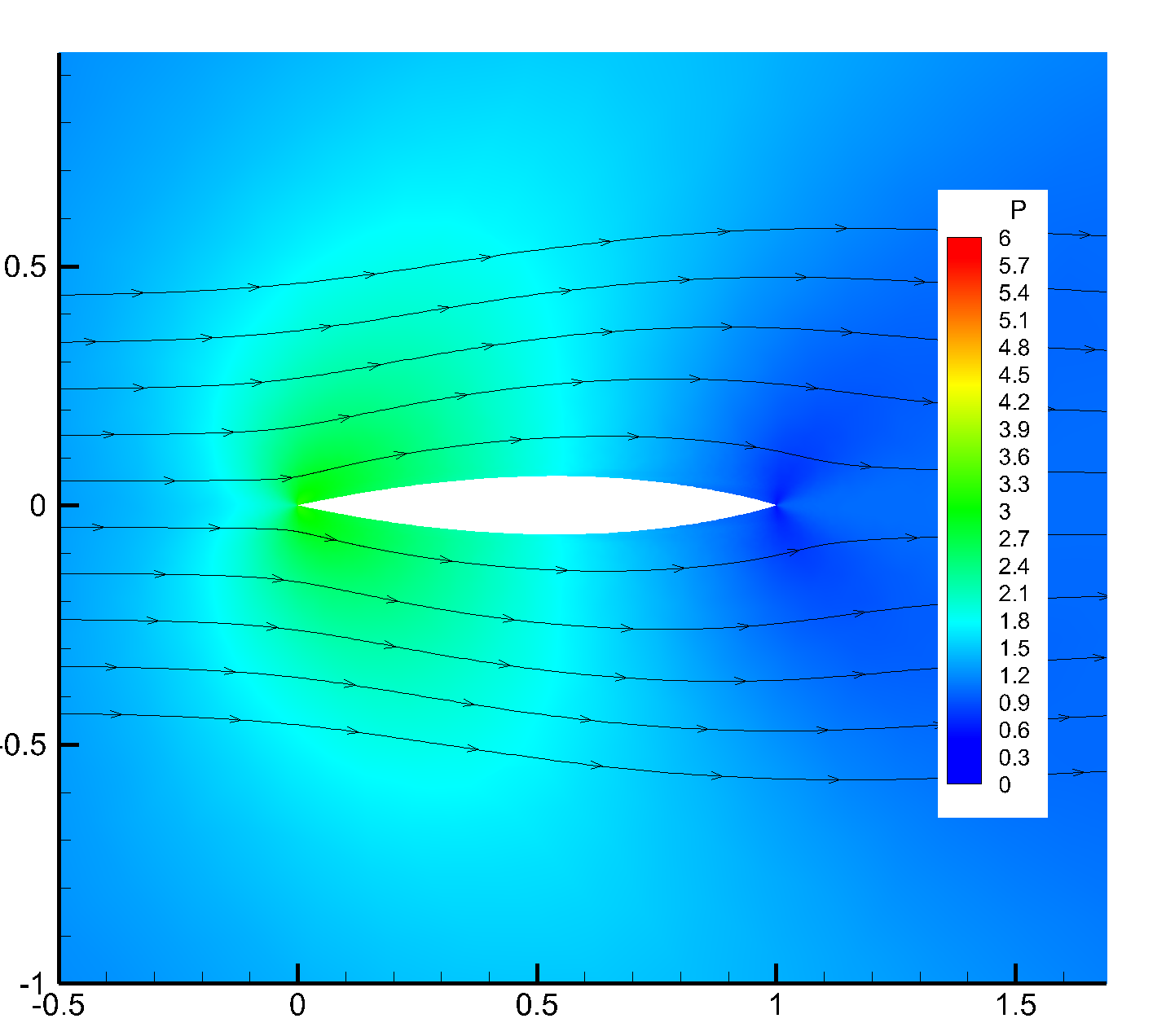}
  \caption{Flowfield comparison at \( \mathrm{Ma}=2 \) between the initial (left column) and optimized (right column) airfoils. The Knudsen numbers are 0.01 and 0.50 in the top and bottom rows, respectively. }
  \label{fig:ma2_flow_compare}
\end{figure}

Figure~\ref{fig:ma2_flow_compare} compares the pressure fields and streamlines around
the initial and optimized airfoils. For \(\mathrm{Kn}=0.01\), the initial airfoil
exhibits a large high-pressure stagnation region near the leading edge and a narrow,
steep pressure gradient along the upper surface. After optimization, the leading-edge
high-pressure region shrinks, isobars in the upper-surface neighborhood become more
widely spaced, and streamlines turn less sharply around the nose and trailing edge,
aligning more closely with the freestream. For \(\mathrm{Kn}=0.50\), the initial airfoil
still produces a relatively broad medium-to-high-pressure region ahead of the leading
edge and noticeable streamline turning around the nose and trailing edge. After
optimization, this elevated-pressure patch contracts, the isobars around the upper
surface become smoother and more evenly spaced, and the low-pressure pocket near the
trailing edge is regularized. In both cases, the optimized shapes weaken excessive
compression and flow turning near the nose, effectively straightening the main flow
passage and reducing drag.

\begin{figure}[t!]
  \centering
  \includegraphics[width=.48\linewidth]{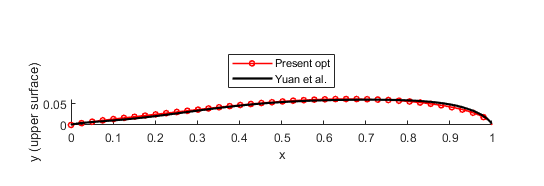}\hfill
  \includegraphics[width=.48\linewidth]{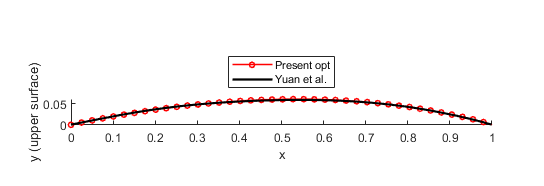}\\
  \includegraphics[width=.48\linewidth]{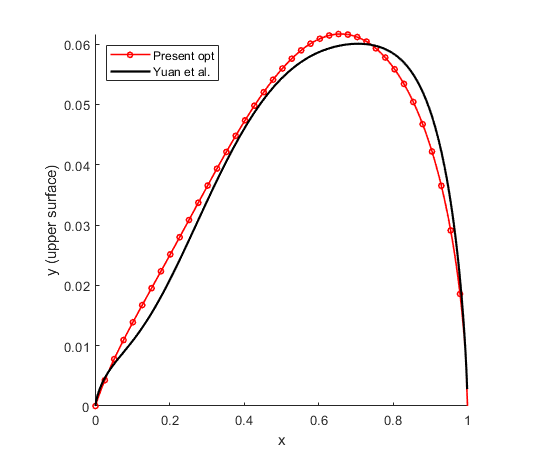}\hfill
  \includegraphics[width=.48\linewidth]{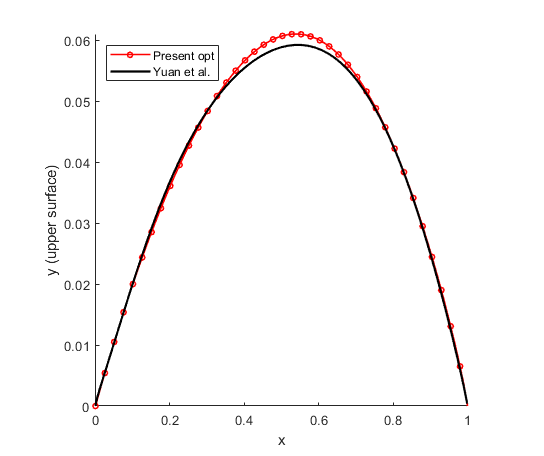}
  \caption{Geometry-only comparison between our optimized thickness distributions and the shape optimization method~\cite{yuan2025adjoint} at \( \mathrm{Ma}=2 \). First row:  physical-scale view
   at \( \mathrm{Kn}=0.01 \) (left) and \( \mathrm{Kn}=0.50 \) (right). Second row: vertically magnified view. 
   }
  \label{fig:ma2_shape_vs_yuan}
\end{figure}

Figure~\ref{fig:ma2_shape_vs_yuan} compares our optimal airfoils against those of the adjoint shape optimization method  \cite{yuan2025adjoint}. For \(\mathrm{Kn}=0.01\), the two \(y_u(x)\) curves almost coincide at the leading and trailing edges; in the main thickness region \(x\approx 0.2\text{--}0.7\), the present optimum lies slightly above that of the adjoint optimization with nearly identical peak location. 
The difference arises from the different kinetic models used: our method employs the modified Rykov kinetic model for polyatomic gases, whereas Ref.~\cite{yuan2025adjoint} uses the BGK model for monoatomic gases. Consequently, the Prandtl numbers differ, leading to differences in the predicted temperature field, which in turn slightly affect the drag calculation. For $\mathrm{Kn}=0.50$, the discrepancy is further reduced: the leading edge, peak, and trailing edge nearly coincide.

Taken together, the above comparisons show that the present optimization framework can automatically yield drag-reducing, geometrically smooth optimal airfoils for \(\mathrm{Ma}=2\) at \(\mathrm{Kn}=0.01\) and \(\mathrm{Kn}=0.50\), while remaining consistent with existing reference results. These findings provide a starting point for extending the optimization
framework to broader sets of operating conditions.

\subsection{Knudsen-number dependence of optimal airfoils at $M_\infty=2$} \label{sec:result-trend}

\begin{figure}[t!]
  \centering
  \includegraphics[width=0.9\linewidth, trim={20 20 30 20},clip]{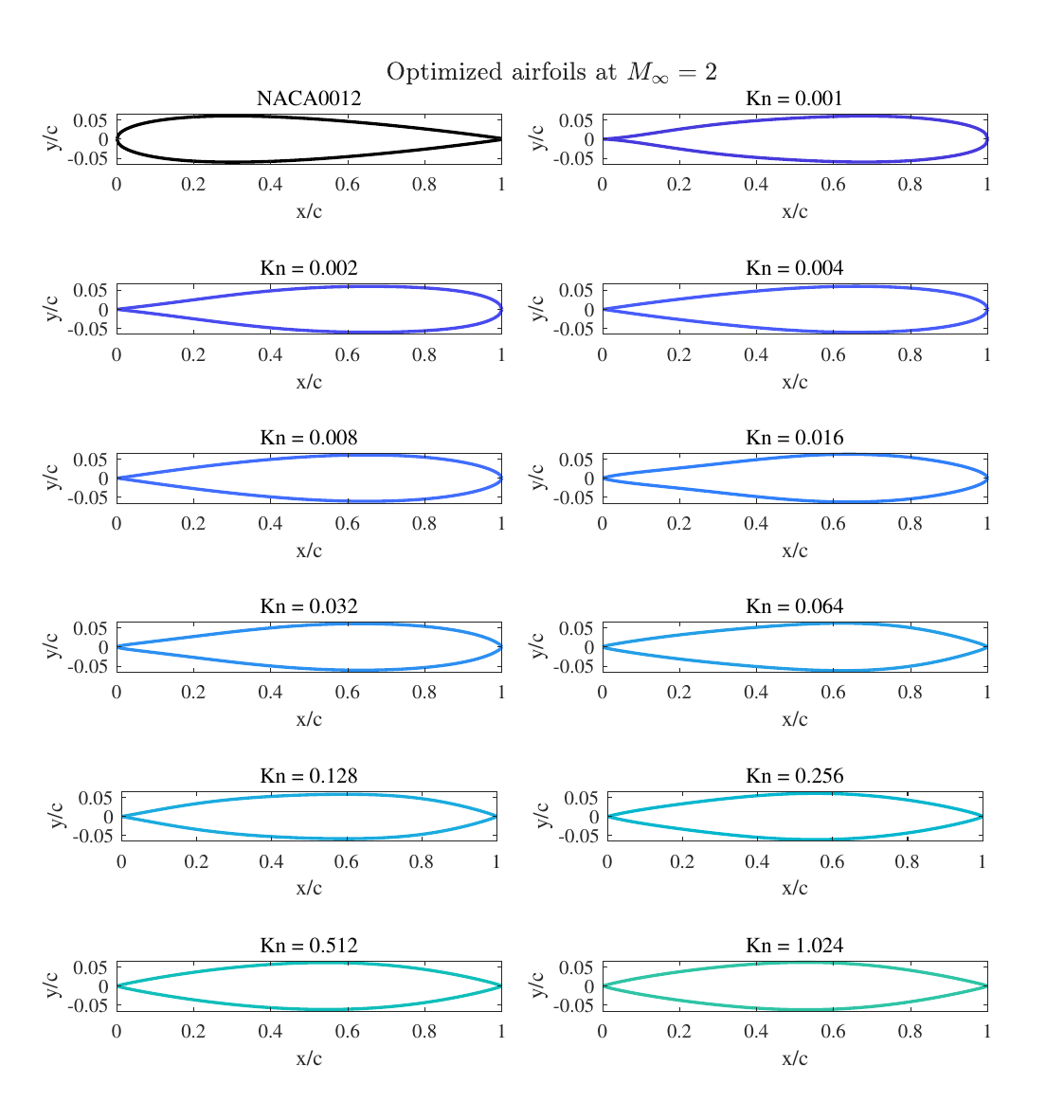}
  \caption{Optimized airfoil shapes at $M_\infty=2$ at 11 different Knudsen numbers. The flow is from left to right.}
  \label{fig:opt_shapes_ma2}
\end{figure}

Figure~\ref{fig:opt_shapes_ma2} shows the optimal airfoils obtained with our
optimization framework at \(M_\infty=2\) over 11 different Knudsen numbers, together
with the initial NACA0012 airfoil. It can be seen that, as
\(\mathrm{Kn}\) varies, the optimal shapes exhibit systematic shifts in peak thickness,
peak-thickness location, and aft loading relative to the baseline section. To quantify
these trends, we examine, as functions of \(\mathrm{Kn}\), the total nondimensional
streamwise force and the corresponding relative drag reduction, the pressure and viscous
contributions to the drag and their respective fractions, as well as several geometric
indicators such as the maximum thickness and its chordwise position.

\begin{figure}[t!]
  \centering
  \includegraphics[width=0.9\linewidth]{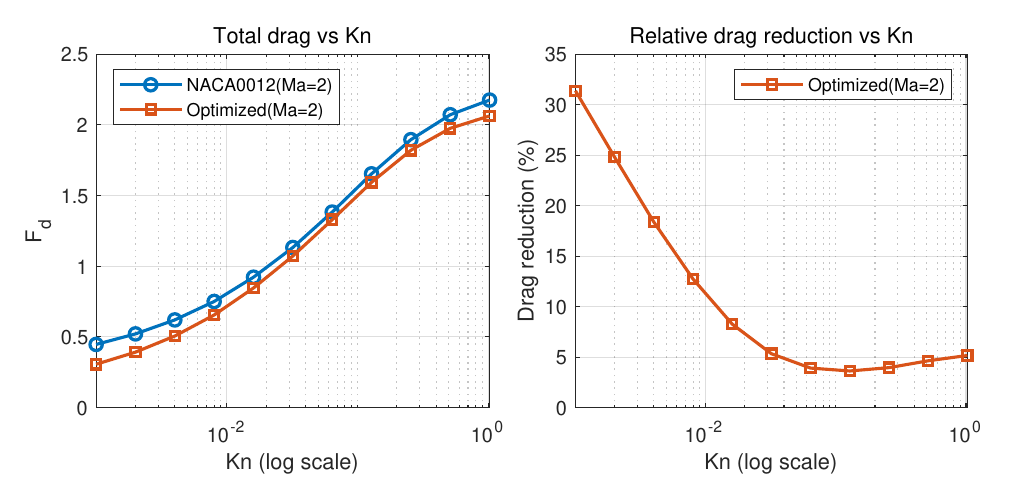}
  \caption{Variation of the total nondimensional streamwise force $F_d$ and the relative drag reduction with Knudsen number for the baseline NACA0012 and the optimized airfoil at $M_\infty=2$.}
  \label{fig:ma2_drag_vs_kn}
\end{figure}

Figure~\ref{fig:ma2_drag_vs_kn} shows that, as $\text{Kn}$ increases, the drag force $F_d$ for both the baseline and optimized airfoils increases monotonically. Although the optimized airfoils consistently exhibit lower drag than the baseline configuration, they follow the same overall trend. Using ``ini'' and ``opt'' to indicate the baseline and optimized airfoils, respectively, the relative drag reduction
\begin{equation}
  \eta_{\text{red}} = \frac{F_d^{\text{ini}} - F_d^{\text{opt}}}{F_d^{\text{ini}}}
  \label{eq:drag_reduction}
\end{equation}
is largest in the weakly rarefied regime, reaching about $30\%$ at the smallest $\text{Kn}$, then decays rapidly with $\text{Kn}$ and levels off near $5\%$ at medium to high $\text{Kn}$. Thus shape optimization yields substantial drag reductions in near-continuum flows but only modest incremental benefits once the flow enters the slip–transition or strongly rarefied regime.

\begin{figure}[t!]
  \centering
  \includegraphics[width=0.9\linewidth]{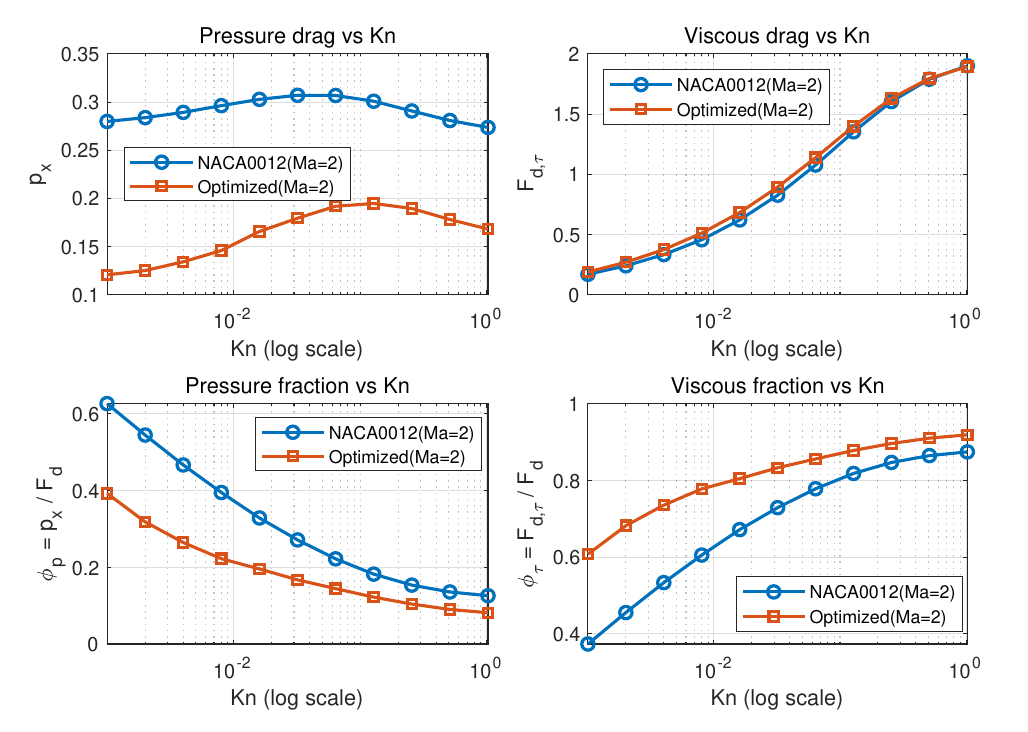}
  \caption{Decomposition of the total streamwise force into pressure drag $p_x$ and viscous drag $F_{d,\tau}$, and the corresponding fractions, as functions of Knudsen number at $M_\infty=2$.}
  \label{fig:ma2_decomp_vs_kn}
\end{figure}

To clarify the origin of this behavior, Fig.~\ref{fig:ma2_decomp_vs_kn} decomposes the total force into pressure drag $p_x$ and viscous drag $F_{d,\tau}$. For the baseline, $p_x^{\text{ini}}$ varies only mildly with $\text{Kn}$ and stays at a roughly constant level, while $F_{d,\tau}^{\text{ini}}$ increases significantly and is mainly responsible for the growth of $F_d$. This is reflected in the pressure fraction $\phi_p$ and viscous fraction $\phi_\tau$:
\begin{equation}\label{eq:phi_p}
  \phi_p = \frac{p_x}{F_d}, \quad
  \phi_\tau = \frac{F_{d,\tau}}{F_d},
\end{equation}
for which $\phi_p^{\text{ini}}\approx 0.6$ at small $\text{Kn}$ and decreases to about $0.1$–$0.2$ at large $\text{Kn}$, whereas $\phi_\tau^{\text{ini}}$ increases from about $0.4$ to nearly 0.9, indicating that the dominant contribution to the drag shifts from pressure to viscous effects. The optimized airfoils reduce $p_x^{\text{opt}}$ by roughly $30$–$50\%$ across the entire $\text{Kn}$ range, while the changes in $F_{d,\tau}^{\text{opt}}$ remain moderate: it is slightly larger than $F_{d,\tau}^{\text{ini}}$ at small $\text{Kn}$, and comparable to or slightly smaller than the baseline at medium and high $\text{Kn}$. Net drag reduction is therefore achieved mainly by weakening the pressure contribution while keeping viscous drag broadly similar. 

\begin{figure}[t!]
  \centering
  \includegraphics[width=0.9\linewidth]{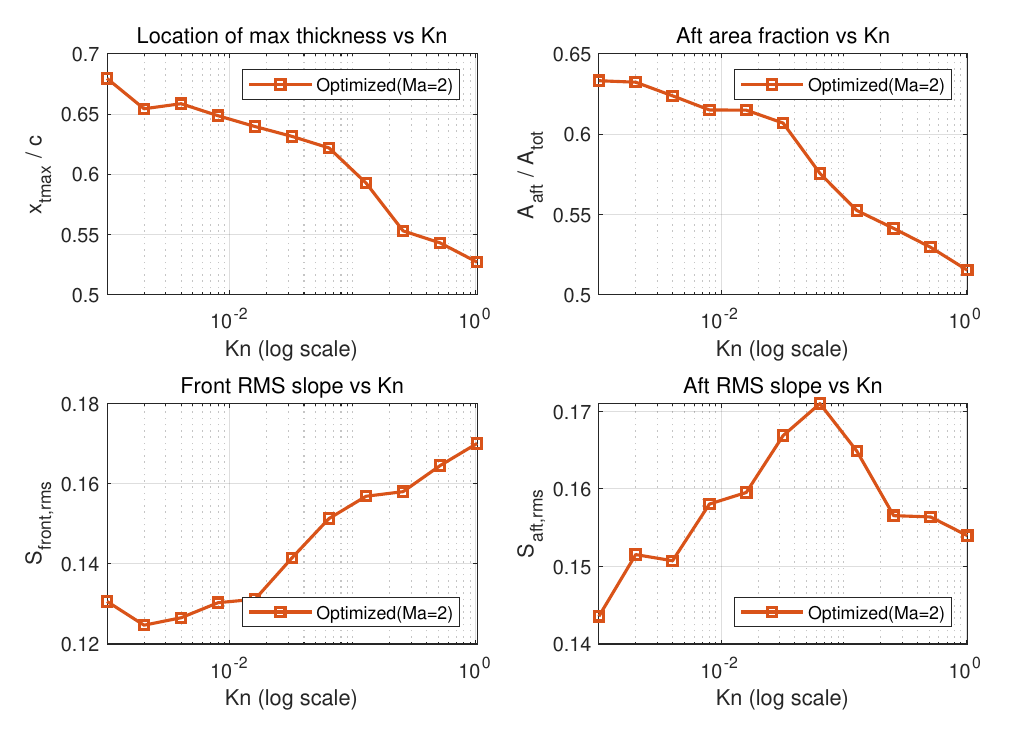}
  \caption{Geometric response of the optimized airfoil at $M_\infty=2$, including the chord-wise location of maximum thickness $x_{t\max}/c$, the after half-chord thickness-area fraction $A_{\text{aft}}/A_{\text{tot}}$, and the RMS slopes in the leading-edge and after-body regions, as functions of Knudsen number.}
  \label{fig:ma2_geom_vs_kn}
\end{figure}

Figure~\ref{fig:ma2_geom_vs_kn} summarizes the geometric response of the optimal
airfoils as \(\mathrm{Kn}\) varies. As \(\mathrm{Kn}\) increases, the chordwise location
of maximum thickness \(x_{t,\max}/c\) moves forward from about \(0.68\) to about
\(0.53\), which helps
alleviate pressure recovery and viscous loading in the rear part. Let
\(A_{\text{tot}}\) denote the total sectional thickness area and \(A_{\text{aft}}\) the
portion of this area downstream of mid-chord \((x/c>0.5)\), it can be seen that the aft half-chord
thickness-area fraction \(A_{\text{aft}}/A_{\text{tot}}\) decreases monotonically
with \(\mathrm{Kn}\), from above \(60\%\) at small \(\mathrm{Kn}\) to about \(50\%\) at
large \(\mathrm{Kn}\). Combined with the forward shift of \(x_{t,\max}/c\), this
indicates that the optimal thickness distribution contracts toward the front as
\(\mathrm{Kn}\) increases, and the aft body becomes relatively thinner.

To track how the nose and tail geometries evolve with \(\mathrm{Kn}\) without entangling
this measure with the shifting location of maximum thickness, we introduce two simple
root-mean-square (RMS) slope indicators on fixed chordwise windows. These windows are chosen empirically
to sample the leading-edge and aft-body neighborhoods (rather than the mid-chord
region where the surface is nearly flat) and to exclude the very last grid points near
the trailing edge where numerical noise can be more pronounced. For a set of chordwise
grid points \(\{x_i\}_{i=1}^N\), the front-region RMS slope \(S_{\text{front,rms}}\)
and aft-region RMS slope \(S_{\text{aft,rms}}\) are then defined as
\begin{equation}
  S_{\text{front,rms}}
  = \sqrt{
      \frac{1}{N_{\text{front}}}
      \sum_{i:\,0 \le x_i/c \le 0.3}
      \left.\left(\frac{dy_u}{dx}\right)^2\right|_{x_i}
    },
  \quad
  S_{\text{aft,rms}}
  = \sqrt{
      \frac{1}{N_{\text{aft}}}
      \sum_{i:\,0.6 \le x_i/c \le 0.98}
      \left.\left(\frac{dy_u}{dx}\right)^2\right|_{x_i}
   }, 
\end{equation}
where \(y_u(x)\) is the upper-surface ordinate and \(N_{\text{front}}\) and
\(N_{\text{aft}}\) are the numbers of grid points in these two windows. These RMS slopes
serve as simple indicators of local surface steepness and small-scale waviness in the
leading-edge and aft-body regions. It can be seen from Fig~\ref{fig:ma2_geom_vs_kn} that
\(S_{\text{front,rms}}\) increases mildly with \(\mathrm{Kn}\) (from about \(0.13\) to
\(0.17\)), indicating a progressively steeper leading-edge region. In contrast,
\(S_{\text{after,rms}}\) first rises from about \(0.145\) to a peak of roughly \(0.18\)
at intermediate \(\mathrm{Kn}\) and then decreases slightly to around \(0.16\) at
\(\mathrm{Kn}=1\), meaning that the aft section exhibits the steepest thickness decay at intermediate
\(\mathrm{Kn}\).

Overall, Figs.~\ref{fig:ma2_drag_vs_kn}--\ref{fig:ma2_geom_vs_kn} show that, at
\(M_\infty=2\), the optimal airfoils across the considered \(\mathrm{Kn}\) range form a
smooth, single-peaked thickness family and adapt through three coupled trends. First,
the total drag increases and becomes increasingly viscous-dominated as \(\mathrm{Kn}\)
grows. Second, optimization at each operating point primarily reduces pressure drag
while only moderately adjusting viscous drag, with the relative benefit largest at
small \(\mathrm{Kn}\). Third, the optimal airfoils show a gradual evolution
from a clearly aft-loaded shape with a more bulged tail at small \(\mathrm{Kn}\)
toward a family where the thickness peak moves forward and the aft section is thinner
and more front-loaded.

\subsection{Mach-number dependence: optimal airfoils at $M_\infty=4$}

\begin{figure}[t!]
  \centering
  \includegraphics[width=0.9\linewidth, trim={20 20 30 20},clip]{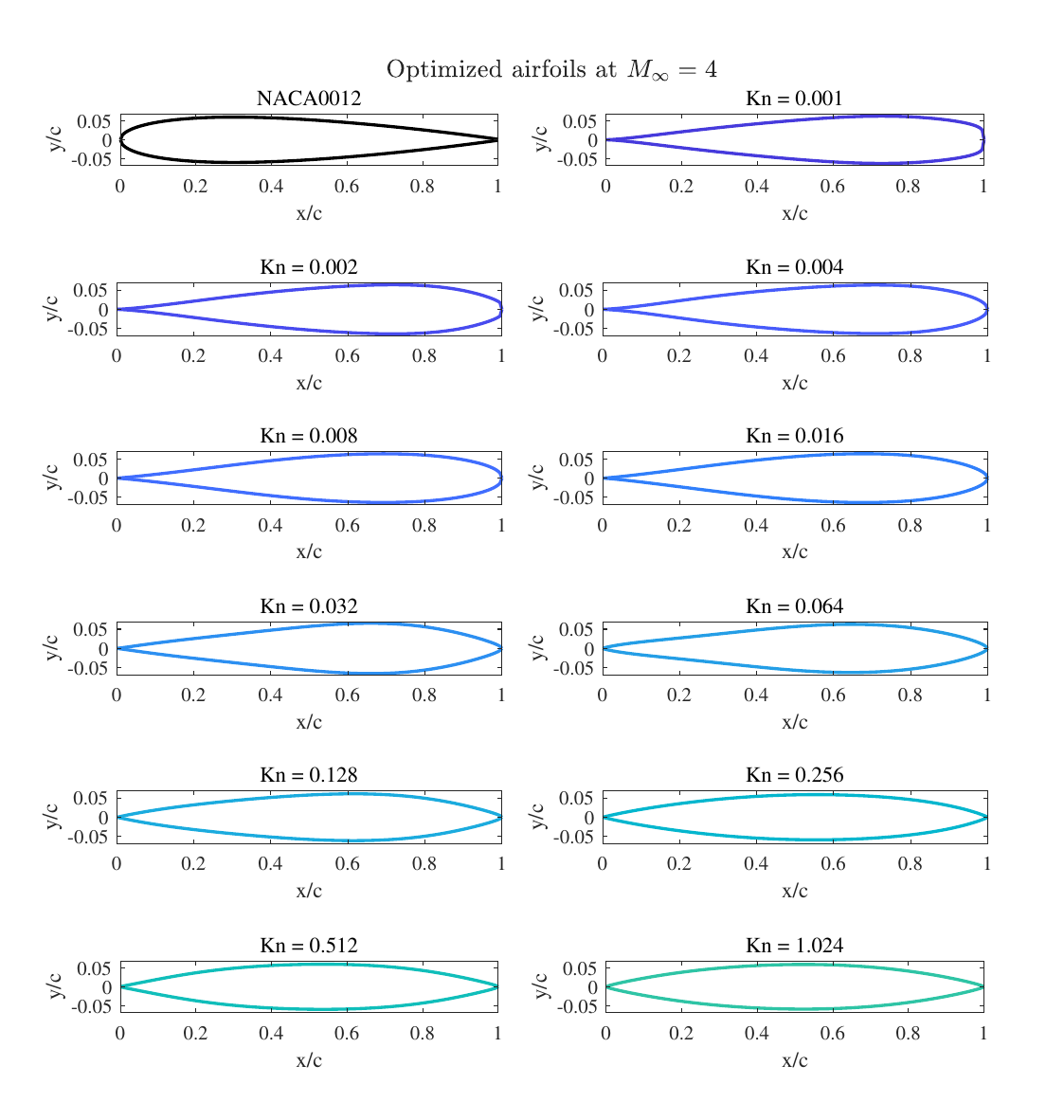}
  \caption{Optimized airfoil shapes at $M_\infty=4$ at 11 different Knudsen numbers, plotted with the same scaling as Figure~\ref{fig:opt_shapes_ma2}.}
  \label{fig:opt_shapes_ma4}
\end{figure}

Figure~\ref{fig:opt_shapes_ma4} shows the optimal airfoils obtained with our
optimization framework at \(M_\infty=4\) over 11 different Knudsen numbers, whose geometric evolution with $\mathrm{Kn}$ broadly mirrors the $M_\infty=2$ case but also exhibits some differences.

\begin{figure}[t!]
  \centering
  \includegraphics[width=0.9\linewidth]{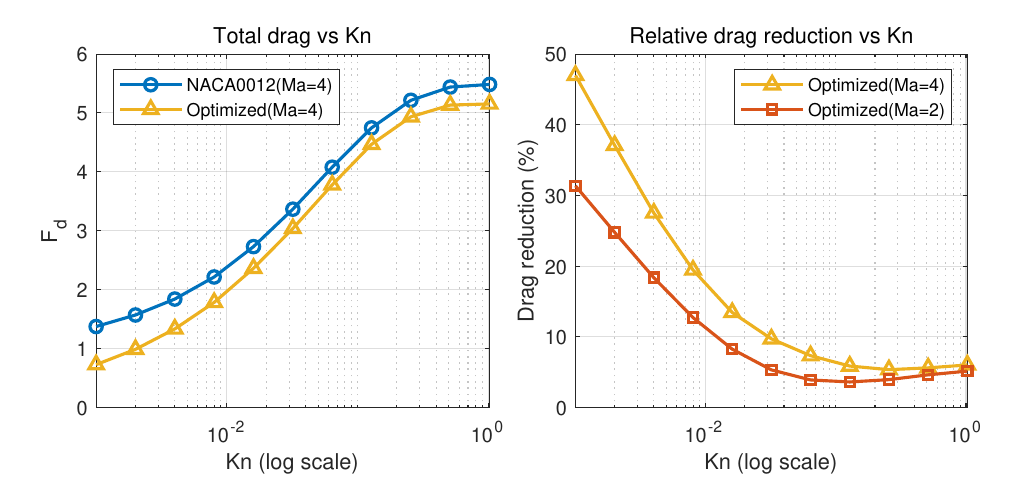}
  \caption{Variation of the total nondimensional streamwise force $F_d$ and the relative drag reduction with Knudsen number for the baseline NACA0012 and the optimized airfoil at $M_\infty=4$. The right panel also shows the drag-reduction curve of the optimized airfoil at $M_\infty=2$ for comparison.}
  \label{fig:ma4_drag_vs_kn}
\end{figure}

Figure~\ref{fig:ma4_drag_vs_kn} shows that, at $M_\infty=4$, the total drag $F_d$ of both the baseline and optimized airfoils still increases monotonically with $\mathrm{Kn}$, but with substantially larger magnitude than those of $M_\infty=2$ due to stronger shocks and compression. For the optimal airfoils, the relative drag reduction reaches about $40$–$50\%$ at low $\mathrm{Kn}$, noticeably exceeding $\approx 30\%$ at $M_\infty=2$, then drops rapidly and approaches a common level of roughly $5\%$ at medium to high $\mathrm{Kn}$ for both Mach numbers. Thus, in strongly rarefied or slip–transition regimes, the additional room for drag reduction via shape optimization is similarly limited at $M_\infty=2$ and $4$.

\begin{figure}[t!]
  \centering
  \includegraphics[width=0.9\linewidth]{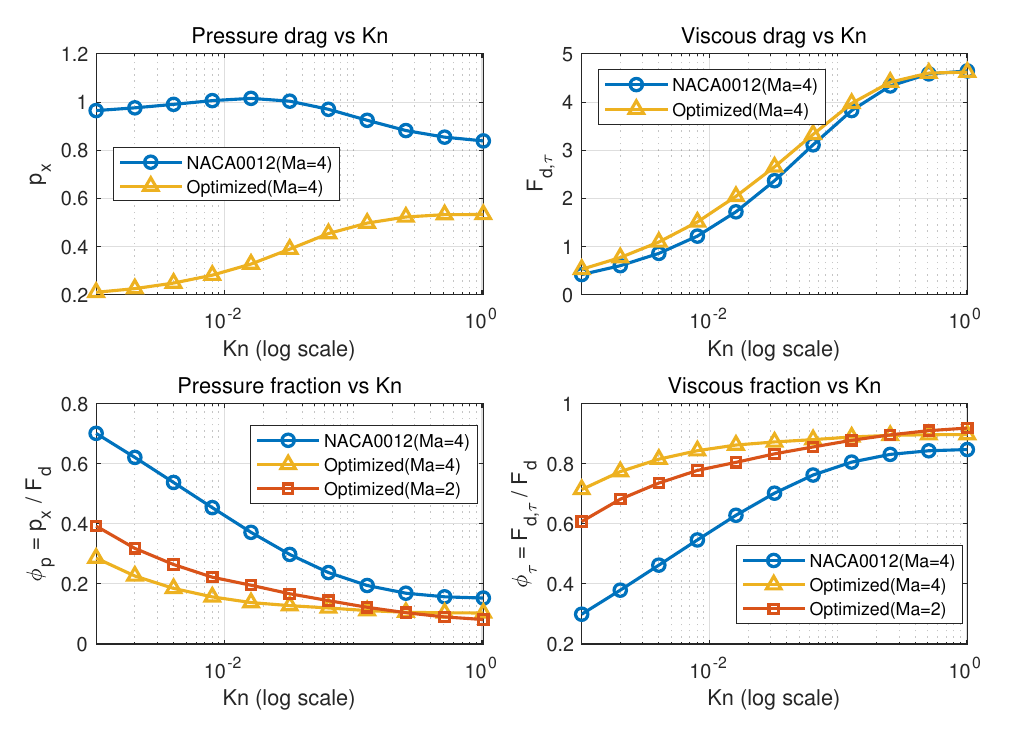}
  \caption{Decomposition of the total streamwise force into pressure drag $p_x$ and viscous drag $F_{d,\tau}$ at $M_\infty=4$, together with the corresponding fractions. The lower panels also include the fractions of the optimized airfoil at $M_\infty=2$ for comparison.}
  \label{fig:ma4_decomp_vs_kn}
\end{figure}

Figure~\ref{fig:ma4_decomp_vs_kn} shows that, for the baseline NACA0012 airfoil at
\(M_\infty=4\), the pressure contribution \(p_x^{\text{ini}}\) stays close to
\(1.0\) over the considered Knudsen numbers, varying only within roughly \(\pm 10\%\)
as \(\mathrm{Kn}\) increases. In contrast, the viscous contribution
\(F_{d,\tau}^{\text{ini}}\) grows rapidly from about \(0.3\) at
\(\mathrm{Kn}=10^{-2}\) to nearly \(5\) at \(\mathrm{Kn}=1\), and is therefore the
main driver of the increase in the total drag \(F_d\), consistent with the
\(M_\infty=2\) case. After optimization, \(p_x^{\text{opt}}\) is clearly reduced at
all \(\mathrm{Kn}\), taking values of about \(0.3\)–\(0.5\), with the largest
relative reduction at low \(\mathrm{Kn}\). The viscous contribution
\(F_{d,\tau}^{\text{opt}}\) is slightly larger than \(F_{d,\tau}^{\text{ini}}\) at
small \(\mathrm{Kn}\), but becomes comparable to or marginally smaller than the
baseline at intermediate and high \(\mathrm{Kn}\). This behaviour again indicates
that the optimization is achieved primarily by weakening the pressure component of the drag.

The lower panels of Fig.~\ref{fig:ma4_decomp_vs_kn} compare the pressure fraction $\phi_p=p_x/F_d$ and viscous fraction $\phi_\tau=F_{d,\tau}/F_d$ across Mach numbers. For the baseline at $M_\infty=4$, $\phi_p^{\text{ini}}\approx 0.6$–$0.7$ at small $\mathrm{Kn}$ and decreases to about $0.1$–$0.2$ at large $\mathrm{Kn}$, indicating that, as rarefaction strengthens, the drag again shifts from being predominantly pressure-driven to predominantly viscous-driven, consistent with the
\(M_\infty=2\) case. Compared with the baseline, the optimized airfoils at both Mach numbers exhibit reduced $\phi_p^{\text{opt}}$ and increased $\phi_\tau^{\text{opt}}$. In most of the $\mathrm{Kn}$ range, the optimized design at $M_\infty=2$ attains a lower $\phi_p$ and higher $\phi_\tau$ than the optimized design at $M_\infty=4$, while the latter lies between the $M_\infty=2$ optimum and the $M_\infty=4$ baseline. Because the baseline pressure contribution is larger at \(M_\infty=4\), there is more room to redistribute drag away from the pressure component, which helps explain the stronger relative drag reduction achieved at low \(\mathrm{Kn}\).

\begin{figure}[t!]
  \centering
  \includegraphics[width=0.9\linewidth]{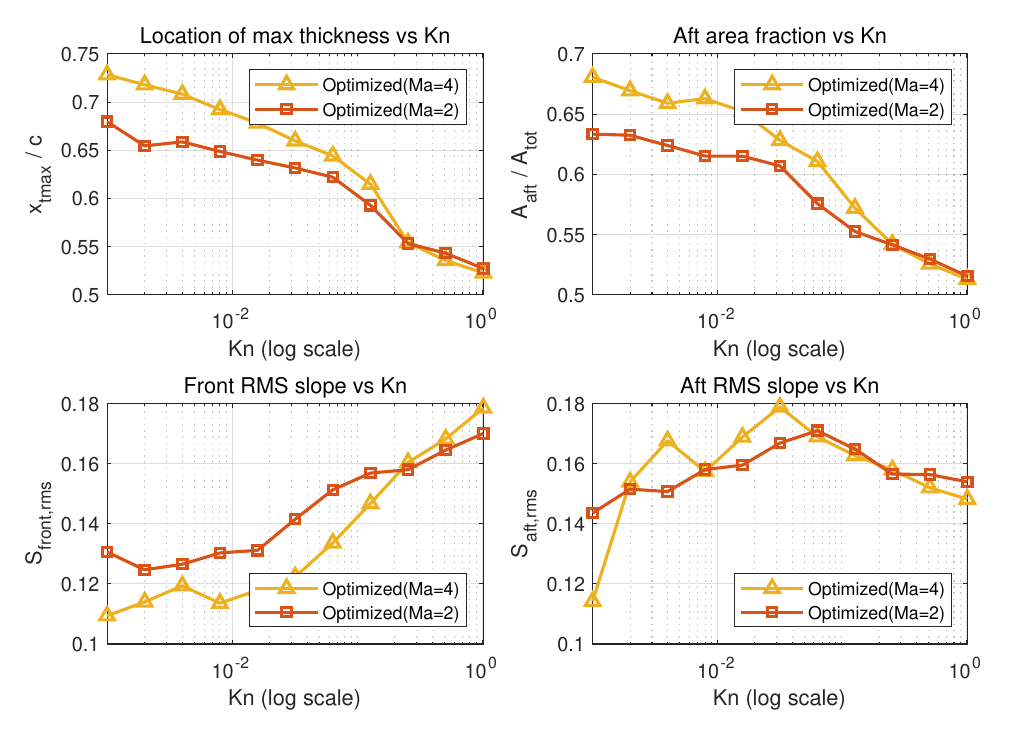}
  \caption{Geometric response of the optimized airfoils at $M_\infty=2$ and $M_\infty=4$ as functions of Knudsen number, including the chord-wise location of maximum thickness $x_{t\max}/c$, the after half-chord thickness-area fraction $A_{\text{aft}}/A_{\text{tot}}$, and the RMS slopes in the leading-edge and after-body regions.}
  \label{fig:ma4_geom_vs_kn}
\end{figure}

Figure~\ref{fig:ma4_geom_vs_kn} summarizes the geometric response of the optimal airfoils at $M_\infty=2$ and $M_\infty=4$ as $\mathrm{Kn}$ varies. At both Mach numbers, the maximum-thickness location $x_{t\max}/c$ shifts smoothly from a clearly after-biased position toward mid-chord with increasing $\mathrm{Kn}$. At small $\mathrm{Kn}$ the peak at $M_\infty=4$ lies further aft than at $M_\infty=2$, and its forward migration is more pronounced; in the high-$\mathrm{Kn}$ regime the two peak locations become close, with $M_\infty=4$ still slightly trailing. Consistently, the after half-chord thickness-area fraction $A_{\text{aft}}/A_{\text{tot}}$ is higher at $M_\infty=4$ than at $M_\infty=2$ at low $\mathrm{Kn}$, reflecting a more strongly after-loaded distribution that helps alleviate leading-edge shocks, and decreases toward similar levels as $\mathrm{Kn}$ increases.

The front- and aft-region RMS slopes further clarify the local geometric trends.
As \(\mathrm{Kn}\) increases, the front-region RMS slope \(S_{\text{front,rms}}\)
grows steadily at both Mach numbers, indicating a progressively steeper increase in thickness from the leading edge toward
mid-chord, with slightly larger values reached
at \(M_\infty=4\). The aft-region RMS slope \(S_{\text{aft,rms}}\) rises with
\(\mathrm{Kn}\) and then decreases mildly at the largest \(\mathrm{Kn}\) for both
cases, but its variation is more pronounced at \(M_\infty=4\),
indicating a stronger thickness decay and more intensive aft-body shaping to
accommodate the stronger shocks and rarefaction effects at higher Mach number.

Taken together with the forward shift of \(x_{t,\max}/c\) and the reduction in
\(A_{\text{aft}}/A_{\text{tot}}\), these trends suggest that, while both Mach numbers
share a smooth, single-peaked thickness family, the optimal designs at
\(M_\infty=4\) rely more heavily on aft-loaded thickness at low \(\mathrm{Kn}\) and on
a stronger forward redistribution of thickness and aft-body steepness as
\(\mathrm{Kn}\) increases, consistent with the pressure-to-viscous rebalancing seen in
the drag decomposition.

\section{Conclusions}\label{sec:conclusion}

In summary, we have developed a solver-in-the-loop shape-optimization framework for
symmetric, thickness-only airfoils operating in rarefied flows. The framework combines
a regularized CST parameterization with a surrogate-assisted Bayesian optimizer and the
efficient kinetic solver GSIS. This combination enables the systematic identification
of drag-minimizing airfoil designs over a range of rarefied-flow conditions under a
limited evaluation budget, while simple geometric filters and admissibility constraints
help exclude nonphysical shapes and maintain numerical robustness.

The results at \(M_\infty=2\) and \(M_\infty=4\) show that, as the flow transitions
from the near-continuum to the strongly rarefied regime, the total nondimensional
streamwise force grows markedly and the drag changes from being pressure-dominated to
viscous-dominated. Shape optimization consistently yields a net drag reduction across
all \(\mathrm{Kn}\) and both Mach numbers. In the weakly rarefied regime, the relative
reduction reaches about \(30\%\) at \(M_\infty=2\) and up to roughly \(40\)–\(50\%\) at
\(M_\infty=4\), reflecting the larger penalty of shock-related pressure drag at higher
Mach number. As the flow enters the slip–transition and strongly rarefied regimes, the
achievable reduction decreases to a few percent and becomes comparable at the two Mach
numbers. A decomposition of the optimized designs shows that, in all cases, the drag
reduction is achieved primarily by substantially reducing the pressure contribution,
while keeping the viscous drag nearly unchanged or only modestly altered. Consequently,
at fixed \(M_\infty\) the optimal airfoils systematically shift the aerodynamic load
from the pressure component toward the viscous component while still lowering the total
drag, and the stronger shock system at \(M_\infty=4\) enlarges the initial pressure
contribution and thus the scope for such pressure-to-viscous rebalancing, explaining
the larger relative gains observed in the weakly rarefied regime.

The geometric response of the optimal airfoil family provides a consistent picture
across both Mach numbers. Over the considered Knudsen numbers, the optimal airfoils
retain a smooth, single-peaked thickness profile, but show a systematic forward shift
of the maximum-thickness location and a gradual transfer of thickness area from the aft
half-chord toward the mid-chord region. At low \(\mathrm{Kn}\), the optimal airfoils
are clearly aft-loaded: the thickness peak lies downstream of mid-chord and more
than half of the thickness area is located in the aft half-chord, especially at
\(M_\infty=4\), where a more aft-loaded distribution helps to accommodate the stronger
shock system. As \(\mathrm{Kn}\) increases, the thickness peak moves smoothly toward
mid-chord and the aft half-chord carries a smaller fraction of the total thickness
area, so that the chordwise load distribution becomes more front-loaded. At the same
time, the thickness variation in the leading-edge region becomes progressively steeper,
whereas in the aft section it is most pronounced at intermediate \(\mathrm{Kn}\).

Taken together, these findings indicate that, over the explored
\((M_\infty,\mathrm{Kn})\) range, the optimal airfoils form a coherent,
rarefaction-aware shape family. They maintain a regularized, single-peaked thickness
profile, reduce the total drag mainly by weakening the pressure contribution while only
modestly changing the viscous contribution, and respond to changes in both Mach and
Knudsen numbers through smooth and interpretable geometric shifts in peak location,
thickness distribution, and local surface steepness.

Beyond the specific numerical values, the study demonstrates that a tightly integrated
workflow—combining CST-based regularization, a carefully constrained admissible set, a
Gaussian-process surrogate with expected-improvement acquisition, and a GSIS kinetic
solver—can robustly handle regimes in which neither classical Navier–Stokes models nor
purely stochastic particle methods are individually efficient, and offers a practical
basis for extending the present approach to cambered sections, three-dimensional wings,
and multi-objective formulations in rarefied aerodynamic design.

\section*{Declaration of competing interest}
The authors declare that they have no known competing financial interests or personal relationships that could have appeared to influence the work reported in this paper.

\section*{Acknowledgments}
This work is supported by the National Natural Science Foundation of China (12402388). 
The authors acknowledge the computing resources from the Center for Computational Science and Engineering at the Southern University of Science and Technology.

\bibliographystyle{unsrtnat}
\bibliography{refs}

\end{document}